\documentclass{aastex}
\usepackage{emulateapj5,danonecolfloat,lscape}

\lefthead{Zirm, Dickinson \& Dey}
\righthead{Massive Ellipticals at $\mathbf z \approx 1$}

\submitted{Received 2002 June; Accepted 2002 November}

\newenvironment{inlinefigure}{
\def\@captype{figure}
\noindent\begin{minipage}{0.999\linewidth}\begin{center}}
{\end{center}\end{minipage}\smallskip}

\def\spose#1{\hbox to 0pt{#1\hss}}
\def\HST{{\it HST}}
\def\lya{\ifmmode {\rm\,Ly\alpha}\else ${\rm\,Ly\alpha}$\fi}
\def\msun{\ifmmode {\rm\,M_\odot}\else ${\rm\,M_\odot}$\fi}
\def\Msun{\ifmmode {\rm\,M_\odot}\else ${\rm\,M_\odot}$\fi}
\def\lsun{\ifmmode {\rm\,L_\odot}\else ${\rm\,L_\odot}$\fi}
\def\Lsun{\ifmmode {\rm\,L_\odot}\else ${\rm\,L_\odot}$\fi}
\def\rsun{\ifmmode {\rm\,R_\odot}\else ${\rm\,R_\odot}$\fi}
\def\Rsun{\ifmmode {\rm\,R_\odot}\else ${\rm\,R_\odot}$\fi}

\def\deg{\ifmmode {^{\circ}}\else {$^\circ$}\fi}
\def\degr{\ifmmode {^{\circ}}\else {$^\circ$}\fi}
\def\degs{\ifmmode {^{\circ}}\else {$^\circ$}\fi}

\def\arcsec{\ifmmode {^{\prime\prime}}\else $^{\prime\prime}$\fi}
\def\asec{\ifmmode {^{\prime\prime}}\else $^{\prime\prime}$\fi}
\def\arcmin{\ifmmode {^{\prime}}\else $^{\prime}$\fi}
\def\amin{\ifmmode {^{\prime}}\else $^{\prime}$\fi}
\def\farcs{\ifmmode \rlap.{^{\prime\prime}}\else
    $\rlap.{^{\prime\prime}}$\fi}
\def\simlt{\mathrel{\spose{\lower 3pt\hbox{$\mathchar"218$}}
     \raise 2.0pt\hbox{$\mathchar"13C$}}}
\def\simgt{\mathrel{\spose{\lower 3pt\hbox{$\mathchar"218$}}
     \raise 2.0pt\hbox{$\mathchar"13E$}}}

\def\rq{$r^{1/4}$ }

\def\Kz{$K$-$z$}

\begin{document}

\def\head{
\title{Massive Ellipticals at High Redshift: 
NICMOS\altaffilmark{1} Imaging of $\mathbf z \sim 1$ Radio Galaxies}

\author{Andrew W. Zirm}
\affil{	Center for Astrophysical Sciences,
       	Johns Hopkins University,
       	3400 N. Charles Street,
       	Baltimore, MD 21218}
\email{azirm@pha.jhu.edu}

\author{Mark Dickinson}
\affil{Space Telescope Science Institute,
       3700 San Martin Drive,
       Baltimore, MD 21218}
\email{med@stsci.edu}

\author{Arjun Dey}
\affil{National Optical Astronomy Observatory\altaffilmark{2},
       950 North Cherry Avenue,
       Tucson, AZ 85726}
\email{dey@noao.edu}

\begin{abstract}
We present deep, $\approx 1.6\mu$m, continuum images of eleven high-redshift 
($0.811 < z < 1.875$) 3CR radio galaxies observed
with NICMOS on-board the {\it Hubble Space Telescope}.  
Our NICMOS images probe the rest-frame 
optical light where stars are expected to dominate the galaxy 
luminosity.  The rest-frame ultraviolet light of eight of these galaxies 
demonstrates the well-known ``alignment effect'', with extended and 
often complex morphologies elongated along an axis close to that of 
the FRII radio source.  As has been previously noted from ground-based 
near-infrared
imaging, most of the radio galaxies have rounder, 
more symmetric morphologies at rest-frame optical wavelengths.
Here we show the most direct evidence that in most cases the stellar hosts are normal
elliptical galaxies with \rq\-law light profiles.
For a few galaxies very faint traces 
(less than 4\% of the total $H$-band light) 
of the UV-bright aligned component 
are also visible in the infrared images.  We derive
both the effective radius and surface 
brightness for nine of eleven sample galaxies 
by fitting one- and two-dimensional surface-brightness models to them. 
We compare the high-redshift radio galaxies to 
lower redshift counterparts.  We find
their sizes are similar to those of local FRII radio source hosts and are in 
general larger than other local galaxies.  
The derived host galaxy luminosities are very high and lie at the 
bright end of luminosity functions 
constructed at similar redshifts. This indicates that the 
high-redshift radio galaxies are likely rare, massive sources.  
The galaxies in our sample are also brighter than the rest-frame size--surface-brightness 
locus defined by the low-redshift sources.
Passive evolution roughly aligns the $z \approx 1$ galaxies 
with the low-redshift samples with a slope equal to $4.7$. 
This value falls intermediate between 
the canonical Kormendy relation ($\approx 3.5$) and 
a constant luminosity line ($=5$).
The optical host is sometimes centered on a local 
minimum in the rest-frame UV emission, suggesting
the presence of substantial dust obscuration.  
We also see good evidence of nuclear point sources (no brighter than 
5\% of the total $H$-band light) in three galaxies. 
Overall, our results are consistent with the hypothesis that these galaxies have already formed 
the bulk of their stars at redshifts greater than $z \simgt 2$, and that the 
AGN phenomenon takes place within otherwise normal, perhaps 
passively evolving, galaxies.  

\end{abstract}

\keywords{
galaxies: formation ---
galaxies: active ---
infrared: galaxies
}
}

\twocolumn[\head]
\altaffiltext{1}{Based on observations made with the NASA/ESA Hubble Space Telescope, 
obtained at the Space Telescope Science Institute, which is operated by the 
Association of Universities for Research in Astronomy, Inc., under NASA 
contract NAS 5-26555. These observations are associated with proposal \#7454.}

\altaffiltext{2}{National Optical Astronomy 
Observatory is operated by the Association of Universities
for Research in Astronomy, Inc. (AURA), under cooperative agreement
with the National Science Foundation.}


\section{Introduction}\label{Intro}

A variety of techniques have been used to find 
galaxies at high redshift, but the hosts of powerful radio sources remain 
uniquely interesting as the most luminous and arguably the most massive 
galaxies known at $z > 1$.  
Radio surveys have successfully found many distant galaxies
with bright optical or infrared counterparts which, in addition, 
often have prominent emission-lines.  
As the number of known radio galaxies (RGs) increased, 
it was discovered that selection via radio flux 
was preferentially finding early-type galaxies.  
At low to intermediate redshift, powerful radio sources 
are almost exclusively found in
giant elliptical (gE) host galaxies, which are often situated in moderately 
rich groups or clusters \citep*{MMS64,HL91}.  
Several radio galaxies at high-redshift have also 
been found to inhabit overdense environments; a sign 
that RGs mark deep potential wells.  
These facts suggest that high-redshift radio galaxies 
(HzRGs) are the progenitors of present-day giant galaxies.  
In addition, the recent discovery of strong correlations between galaxy and 
central black hole masses \citep*{MTR+98,GBB+00,FM00} 
implies an evolutionary scenario in which \emph{all} giant galaxies 
at \emph{all} observed redshifts harbor super-massive black-holes which 
undergo (perhaps recurrent) accretion episodes to the present day.  
If not prone to strong selection effects, 
these relatively rare sources make excellent 
discriminators of structure formation and evolution models.  
However, the question remains: how might 
the presence of an AGN (both dynamically and its associated radiation and 
outflows) affect a host galaxy's formation and evolution, particularly 
in the extreme cases such as the 3CR radio sources?

There is evidence that the powerful nuclei of radio galaxies play a major role 
in the stellar hosts' evolution.  
Local radio galaxies frequently show morphological peculiarities such as tidal tails and dust 
lanes \citep[e.g.,][]{HSB+86,SH89} which are usually absent in 
the general galaxy population.  While these features do not usually dominate 
the galaxy light (i.e., they can still be classified as ellipticals), 
they have been interpreted by some to suggest that radio activity is triggered 
by galaxy interactions and mergers \citep*[e.g.,][]{HCB85}.  
At higher redshifts, $z \simgt 0.7$, powerful radio galaxies have 
strong rest-frame UV line and continuum emission which is often extended on an 
axis similar to that of the radio source \citep*{MvBS+87,CMvB87}.  
This ``alignment effect'' has a dominant influence on the measured 
morphological and photometric properties of high-redshift 
radio galaxies, at least at ultraviolet 
wavelengths where the blue, AGN-related continuum emission outshines the stellar 
light from the presumed extant host galaxy.  

Optical observations with 
{\it HST}/WFPC2 have revealed the 
remarkably peculiar and complex morphologies of this 
aligned light \citep*{DDS95,BLR97,MMdK+97}.  
The rest-frame UV continuum is often segregated into several discrete 
clumps which are strung along the radio axis.  
There are three viable explanations for the alignment effect:
scattering of AGN light by dust or electrons, star-formation 
induced by the passage of the radio jets through the 
galactic ISM, nebular continuum emission \citep*{DTS+95}
 or some combination thereof.  
It has been established from the high degree of polarization
that the spatially extended, 
aligned, UV continuum emission in many $z \approx 1-2$ RGs is scattered light 
\citep*[e.g., ][]{JE91,dSACF94,DCvB+96,CDvB+97,TCO+98}. 
Data on the higher redshift RGs ($z>2$) are more limited, but at least 
one galaxy shows evidence of young stars contributing to the aligned light \citep*{DvBV+97}.  
If alignment by induced star-formation is prevalent at 
early times, then a large fraction of the stars in the galaxy 
may have been formed via this process.  

On the other hand, near-infrared 
observations of distant radio galaxies have found a smooth 
magnitude-redshift relation (\Kz) with remarkably small scatter 
out to at least $z \approx 3$ \citep*[e.g., ][]{LL84, ERD+93, MCC93, JRE+01}.  
The relation is well-described by a coeval, passively evolving 
galaxy population which formed at $z \simgt 4$.  
Less luminous radio galaxies \citep*{ERL+97} over the same redshift range 
display a \Kz\ relation analogous to their powerful 3CR counterparts.  
In effect the radio selection has chosen a particular class of galaxy 
from the general population.  This is consistent with the local 
association of radio sources with giant elliptical galaxies.  

To explain the alignment effect and \Kz\ diagram simultaneously, 
previous authors \citep*{RLS+92,BLR98} have adopted 
a two-component source model.  A massive, old, early-type galaxy accounts 
for the \Kz\ relation and a bright, but dynamically insignificant, 
AGN-related contribution (scattered nuclear light and/or 
jet-induced star formation) dominates the rest-frame UV light.
Ground-based infrared observations have supported this 
view (\citealt{RLS+92}; \citealt*{BLR97}), but have been 
limited by the relatively poor angular resolution attainable from 
the ground. However, two-dimensional fitting of 
higher angular resolution  {\it HST}/WFPC2 
imaging of several $z \approx 1$ 3CR radio galaxies 
\citep*{McD00} find elliptical-like host galaxies.
Such direct observations of the elliptical 
component can corroborate this theory and perhaps provide a quantify 
the biases inherent in using AGN hosts to trace galaxy evolution.  

To target the stellar host and its morphology requires three 
observational capabilities: 1) access to rest-frame optical 
or longer wavelengths to increase the contrast between stellar 
light and AGN-related emission, 2) filters which target continuum 
emission and avoid bright emission-lines, and 3) high angular-resolution to discern 
morphological features with clarity.  
{\it HST}/NICMOS currently provides the best technology to address 
these three criteria.  
We therefore obtained {\it HST}/NICMOS imaging of a sample 
of 11 powerful radio galaxies at $0.8 < z < 1.8$.  
Our targets are a representative subsample of the most powerful radio 
galaxies at these redshifts.  Although these 11 galaxies do not 
constitute a statistically complete sample of high redshift radio galaxies, they 
do span the full range of optical and near-infrared (NIR) photometric properties shown by 
this population as a whole.  
For example, the strength of the alignment effect in our sample ranges from essentially 
zero (e.g., 3CR~65) to very well aligned (e.g., 3CR~368 or 3CR~266).  

We present both {\it HST}/NICMOS and {\it HST}/WFPC2 imaging 
from our own proposals as well as from 
the \HST\ data archive.  In the next section we describe the data acquisition 
and reduction.  In \S\ref{Analysis} we describe our profile fitting procedures 
and error analysis. In \S\ref{ResDisc} we
present the derived physical parameters and compare them to local galaxy samples, and 
discuss the implications of our results.  

Throughout this paper we assume $H_0 = 65$ $\textrm{km s}^{-1} \textrm{Mpc}^{-1}$ and 
$(\Omega_{\Lambda},\Omega_{M}) = (0.7,0.3)$.  
With this cosmology $1\arcsec$ subtends 8.6 kpc at $z=1$.  

\section{Observations and Data Reduction\label{Obs}}

The sample galaxies were chosen from the 
3CR radio survey to represent 
powerful radio galaxies at $z \approx 1$.  Our 11 galaxies 
cover $0.8 < z < 1.8$ and exhibit the range 
of host galaxy properties present in the 3CR as a whole.
As with any flux-limited sample, 3CR galaxies show a strong 
radio-luminosity--redshift correlation.  This should not greatly 
affect most of our conclusions, but should be kept in mind 
when comparing radio galaxies over different epochs,
as we do in \S\ref{ResDisc}.  Table~\ref{obstab} lists 
the sample galaxies (Col.~1), their coordinates (Col.~2-5) and 
redshifts (Col.~6) along with the exposure 
parameters, including exposure time (Col.~9), filter (Col.~7), number of dithers 
(Col.~10), rest-frame wavelengths covered (Col.~8), and estimated emission-line contamination 
(Col.~11).

\vspace{2.0truecm}
\subsection{Data Acquisition\label{DataAcq}}

Our NICMOS observations were taken with Camera 2 between 
December 1997 and July 1998.  The NICMOS filters 
F160W ($1.60\mu$m) and F165M ($1.65\mu$m) were used to
obtain images in rest-frame wavelengths ranging from 
approximately 6000 to 9000\AA\ (exact wavelength ranges 
can be found in Table~\ref{obstab}).  The alignment effect 
is especially prominent in the narrow emission line 
gas \citep*[e.g., ][]{McSvB95}.  Therefore, to avoid this contamination of the stellar component in terms 
of both luminosity and morphology, we used the 
known redshifts of the sources and the composite RG spectrum of \citet{MCC93} to 
choose the filter with the least line contamination.  We attempted 
to avoid potentially bright emission lines such as H$\alpha$ and,
for the higher-redshift objects, [OIII]~$\lambda\lambda$~4959,5007.  
In the case of 3CR~256 we took exposures in both F160W and F165M to obtain continuum 
and [OIII]~$\lambda\lambda$~4959,5007 images.  
The resulting images show some contamination which has been estimated by 
using a template spectrum with and without emission lines.  
We note that these estimates assume a spatially-constant equivalent 
width, whereas this value is likely to vary over the area of the galaxy.  
The estimated emission-line 
contamination is listed in column~11 of Table~\ref{obstab}.  
The telescope was dithered between exposures for all the objects, 
with 8-13 dither positions per object, to better sample the 
instrumental PSF, which is already critically sampled at 1.6$\mu$m,
and to aid with the recognition and elimination of data artifacts 
such as dead pixels or ``grot'', hot pixels, the unstable central column 
of NIC 2, and persistent cosmic-ray afterglow.  
The exposures themselves were taken using a combination of 
the MIF512 and MIF1024 NICMOS MULTIACCUM readout sequences, providing exposure times 
of 512 and 1024 seconds, respectively.  
In Figure~\ref{TrioFig} we present the NICMOS images (second panels),
the rest-frame UV images from WFPC2 (first panels), which were registered and 
interpolated to the NICMOS images, and the two-dimensional fit
residuals (third panels), which were created by subtracting the best-fit 
model galaxy from the data as described in \S\ref{2DFit}.

\begin{figure*}[t]
\epsscale{1.9}
\plotone{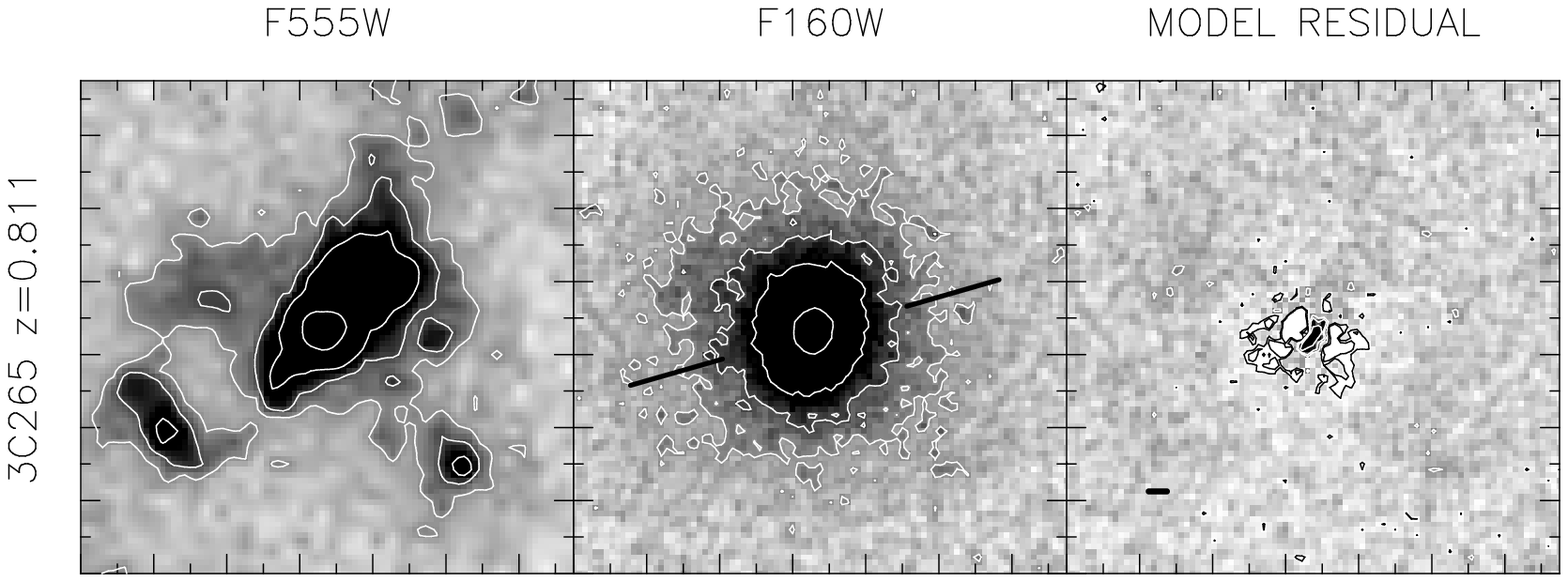}
\end{figure*}

\begin{figure*}
\plotone{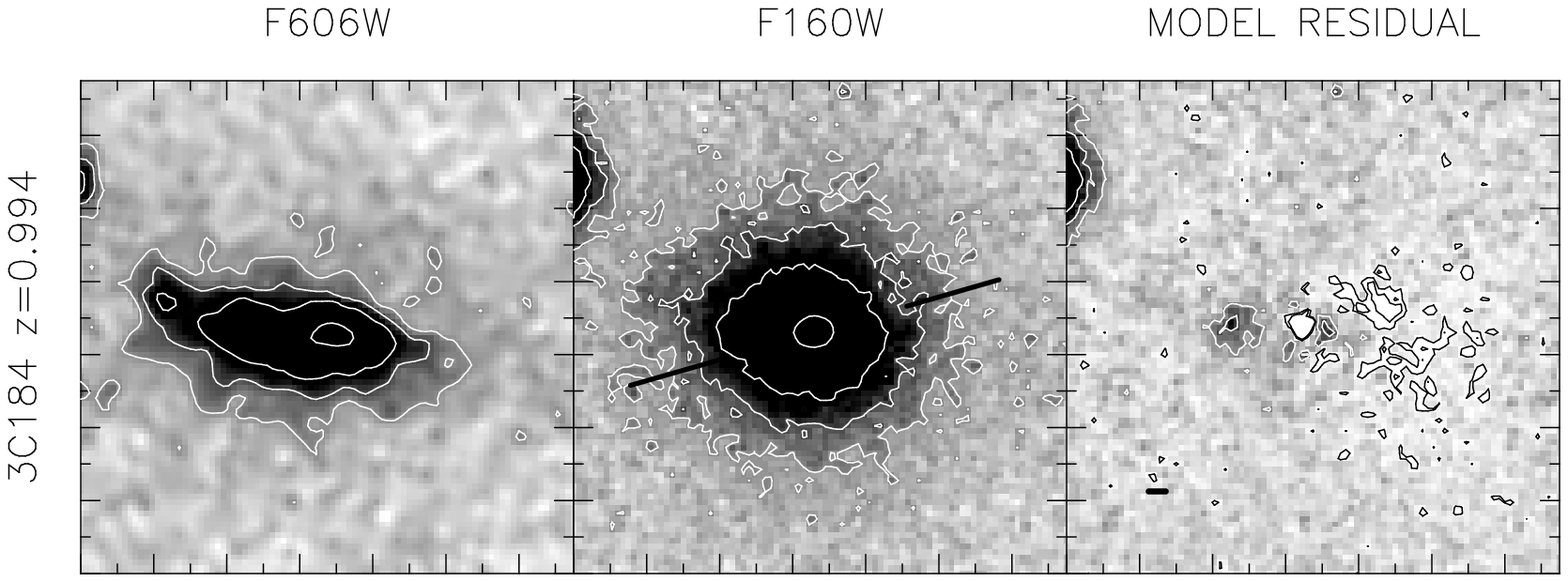}
\end{figure*}

\begin{figure*}
\plotone{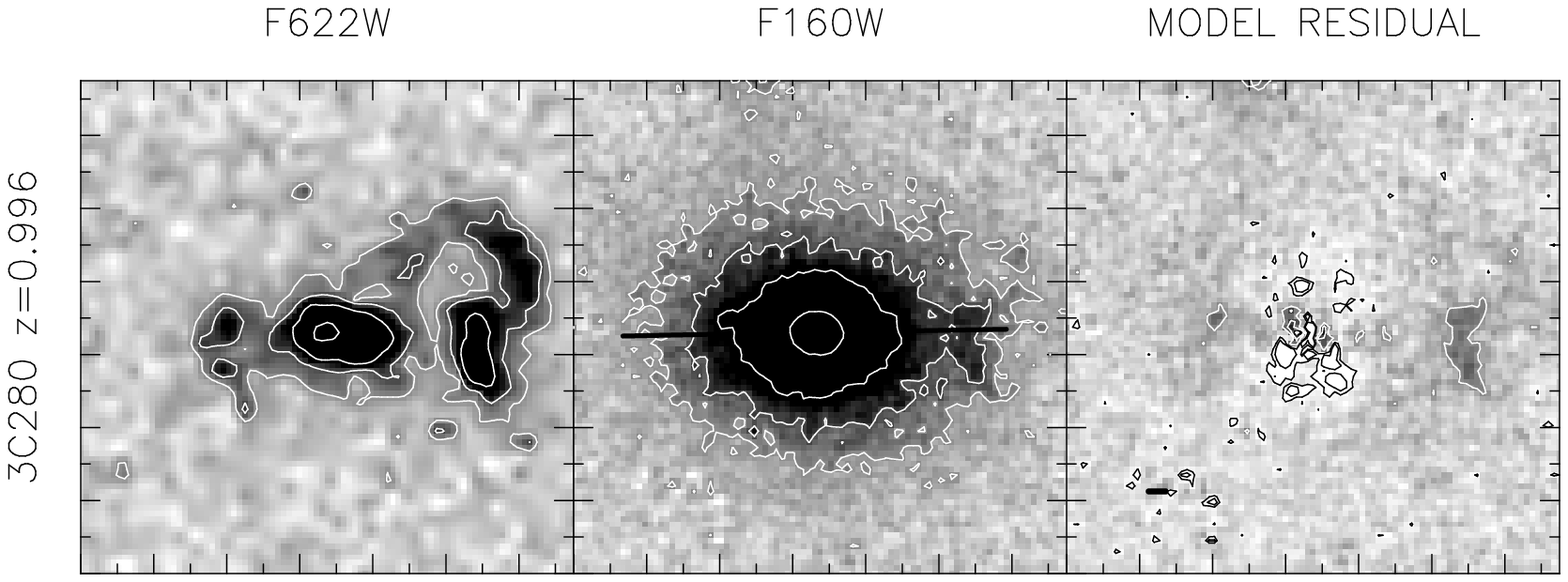}
\end{figure*}

\begin{figure*}
\plotone{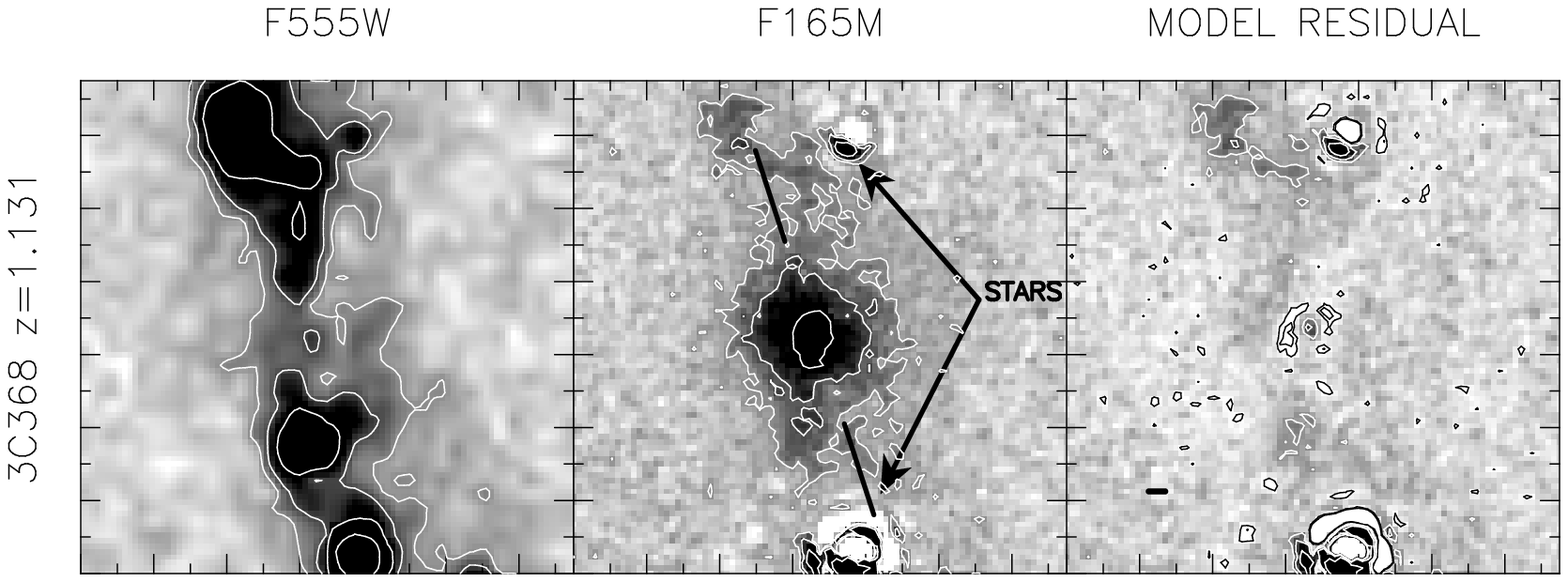}
\end{figure*}

\begin{figure*}
\plotone{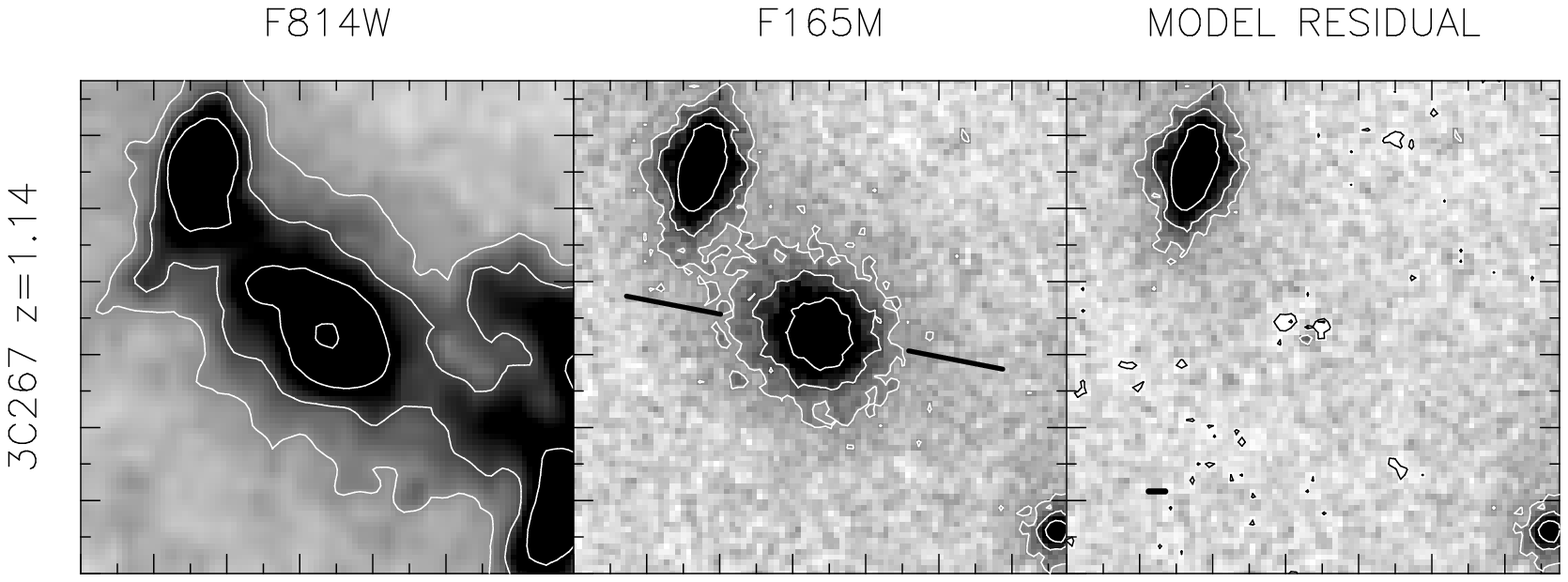}
\end{figure*}

\begin{figure*}
\plotone{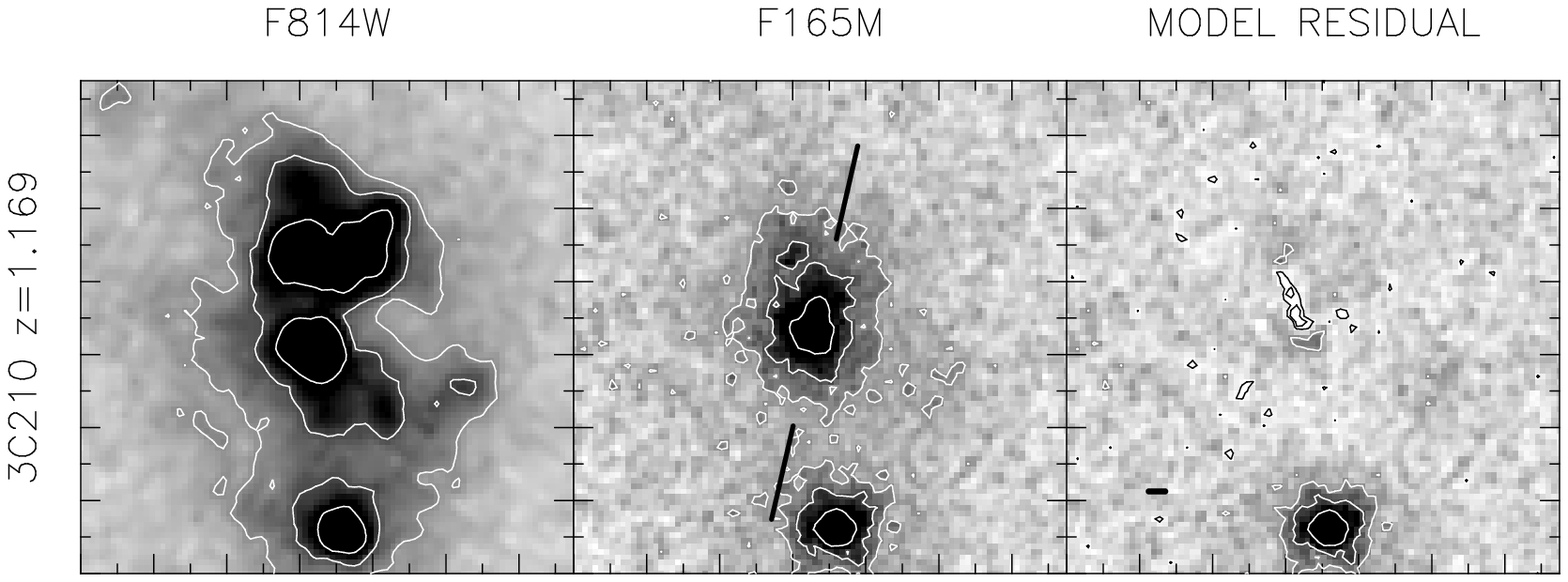}
\end{figure*}

\begin{figure*}
\plotone{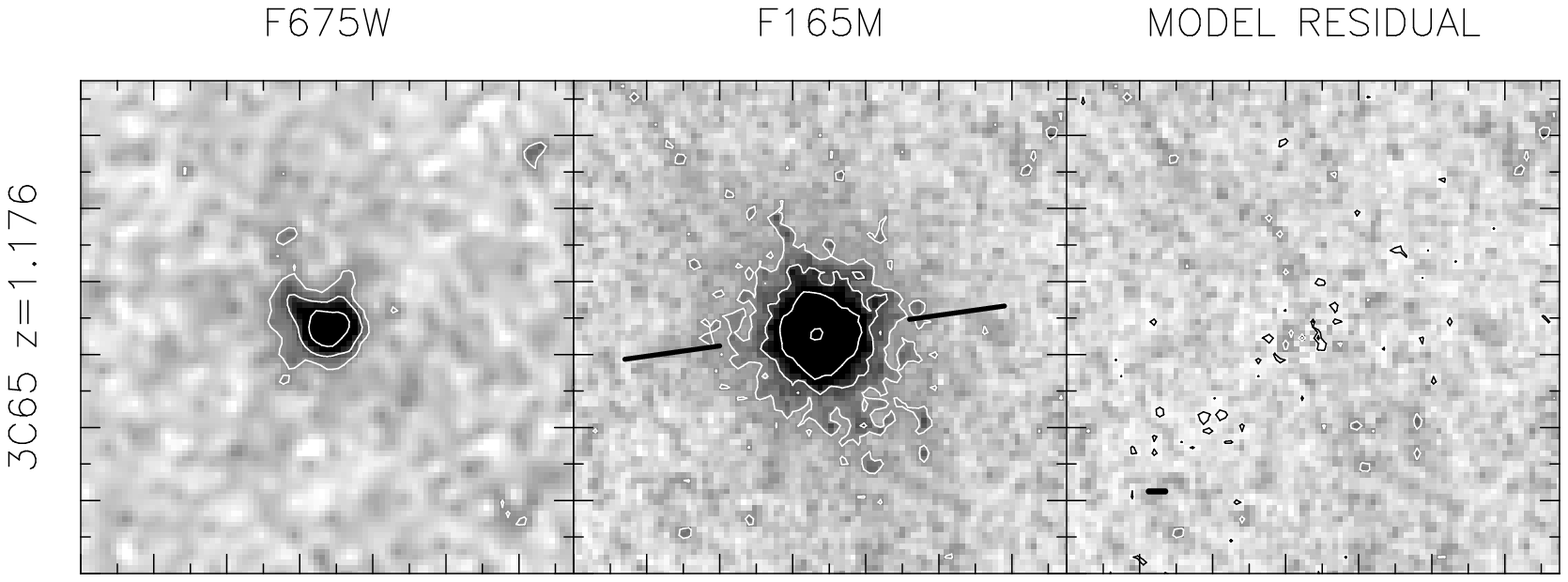}
\end{figure*}

\begin{figure*}
\plotone{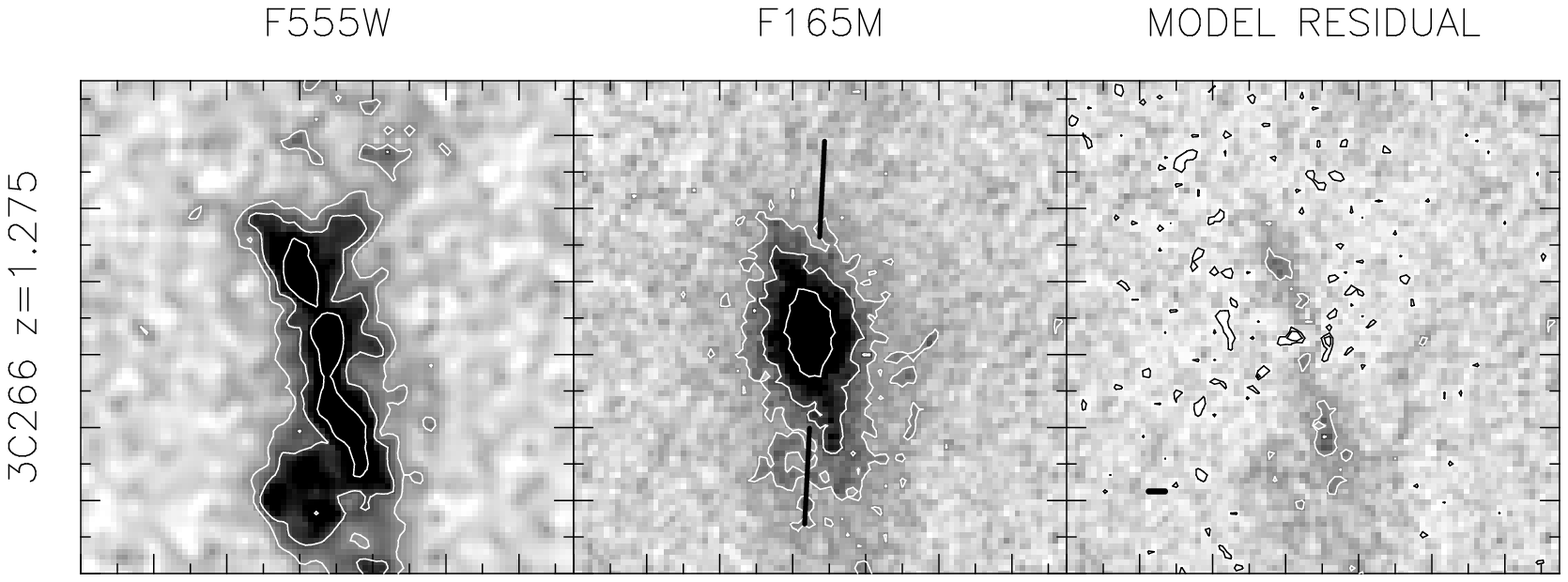}
\end{figure*}

\begin{figure*}
\epsscale{1.266666}
\plotone{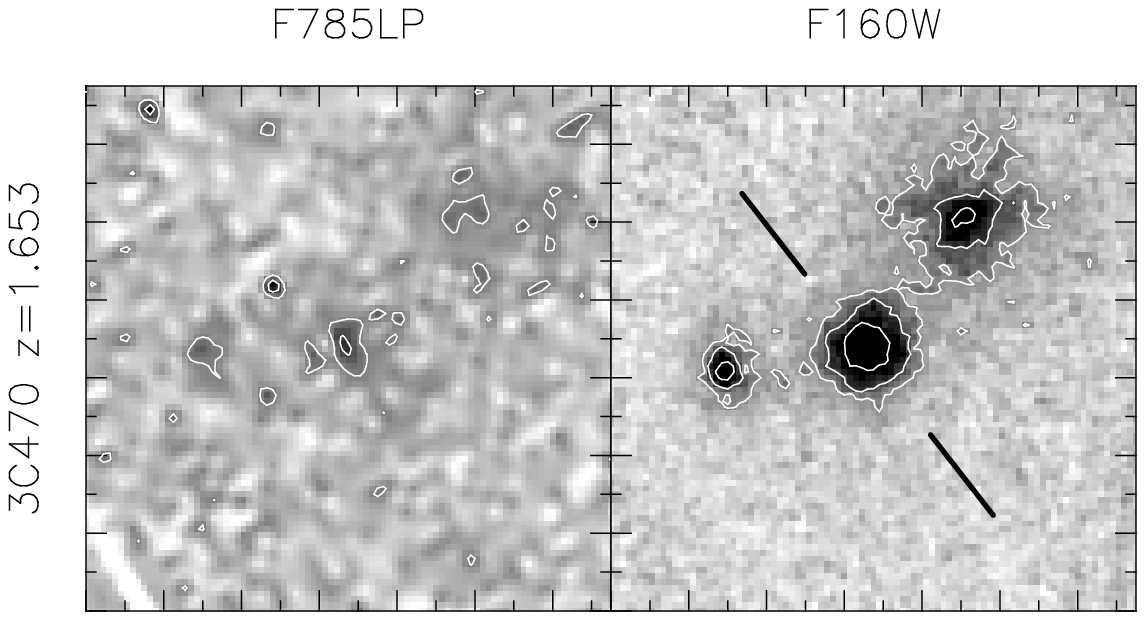}
\end{figure*}

\begin{figure*}
\epsscale{1.9}
\plotone{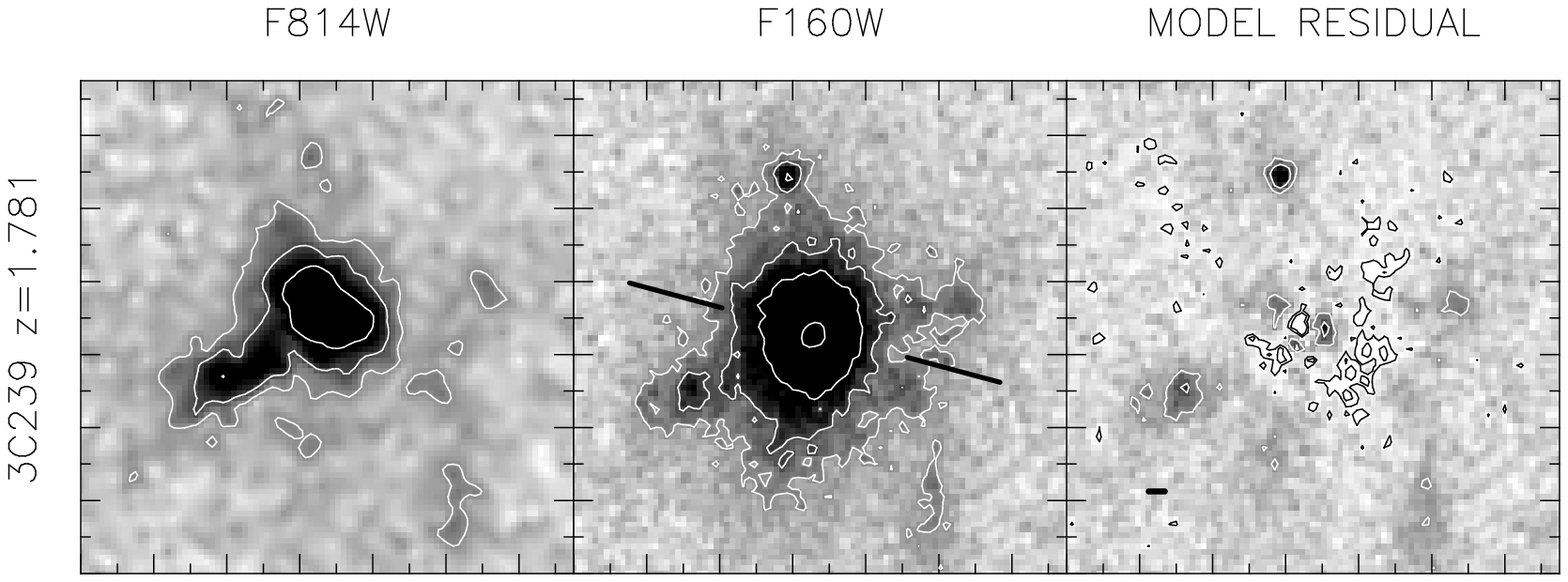}
\end{figure*}

\begin{figure*}
\epsscale{1.266666}
\plotone{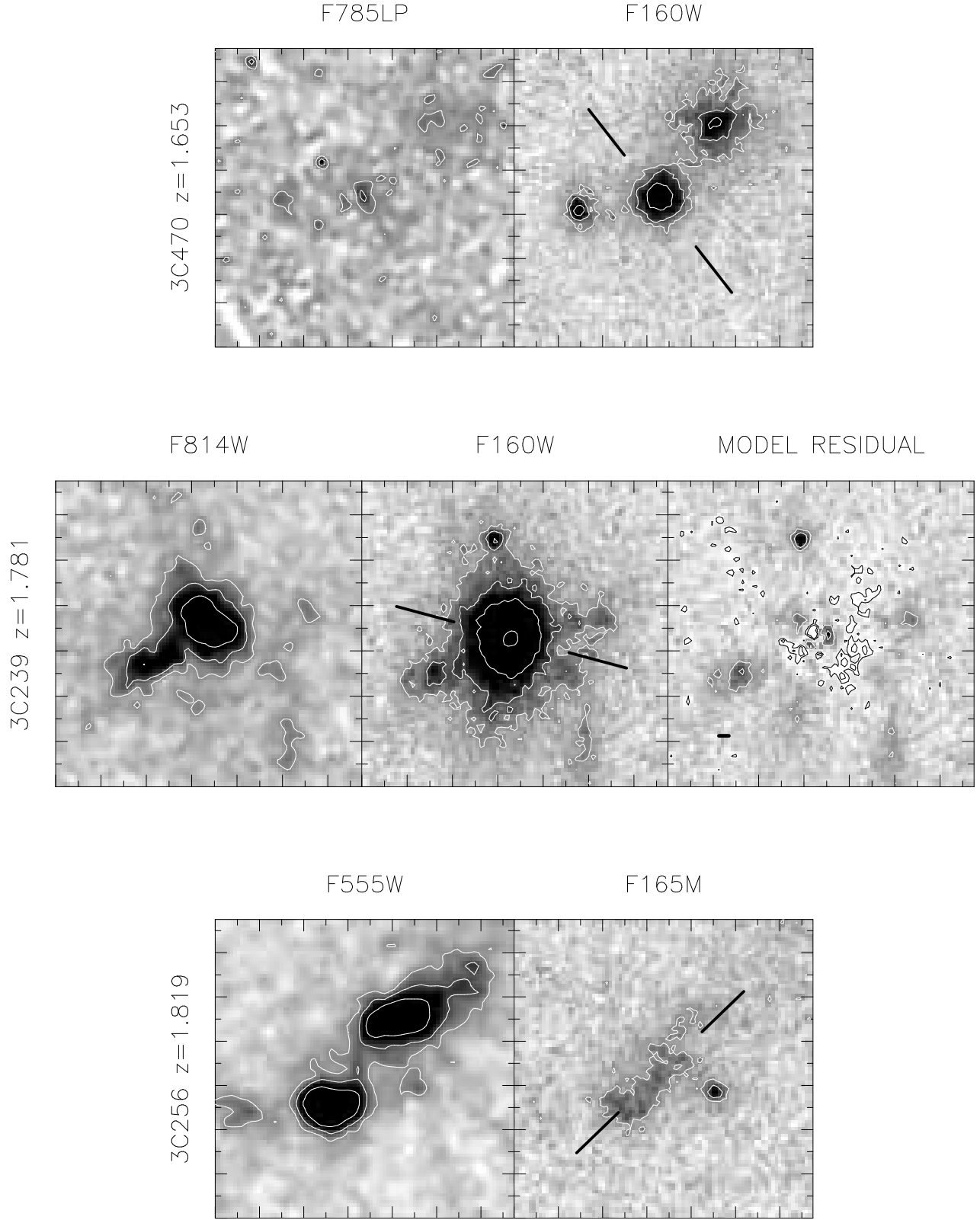}
\caption{WFPC2, NICMOS, and model residual images for our 9 fit galaxies,
presented in ascending redshift order (from 
3CR~265 at $z=0.811$ to 3CR~256 at $z=1.819$).  
The images are shown in the standard North up and East left 
orientation.  
Note the similarity between the NICMOS residuals 
and the rest-frame UV WFPC2 images, suggesting a simple connection 
between the two, i.e., that the NICMOS residuals are simply the 
long-wavelength tail of the light seen in the UV.  The data for 3CR~470 and 256 
are presented despite not being analyzed in the same manner.  The contours 
are at 5, 10, 20 and 100$\sigma$ above (or below) the locally determined background.  The 
major tickmarks are $0\farcs5$ apart and the panels are $3\farcs375$ on a side
\label{TrioFig}
}
\end{figure*}

\subsection{Data Reduction\label{Redux}}

The individual frames were reduced using a combination 
of standard IRAF\footnote{IRAF is distributed by the 
National Optical Astronomy Observatories, 
which are operated by the Association of Universities for Research 
in Astronomy, Inc., under cooperative agreement with the 
National Science Foundation.}/STSDAS pipeline 
routines (CALNICA) and our own techniques.  
The most prominent anomalies in our data were: 
the ``pedestal'', a random DC offset which 
varies by detector quadrant; cosmic-ray persistence, time-decaying 
charge deposits from cosmic-ray hits during {\it HST's} passage through 
the South Atlantic Anomaly; and noise associated with the synthetic dark 
frames in the standard pipeline \citep*{NICDATA}.
The ``pedestal'' subtraction was performed using the PEDSKY task now included 
in STSDAS, but originally written for use with this dataset.  It performs 
an iterative subtraction of both a constant offset per detector quadrant 
and a ``sky'' component which is modeled as a constant (over the whole detector) 
multiplied by the flat-field pattern.  
The solution minimizes the pixel-to-pixel RMS in each quadrant to 
converge on the best fits to these parameters.  

The synthetic darks now used in CALNICA are ``super-darks'' 
generated using many more individual frames than the darks which were 
available when we reduced 
these data.  Also, it is now possible to generate temperature-dependent 
dark frames for individual exposures, which help reduce bias 
gradients in the images (called ``shading'').  
To correct our data for the noise pattern introduced by the 
original darks, we constructed median frames from the sky 
and pedestal subtracted images for each filter and 
readout sequence combination.  We combined both our 
data and those from a similar observing program 
(M. Giavalisco, private communication) using a median, and subtracted 
the resulting image from our data.  We were able to reduce the amount of pixel-to-pixel 
noise introduced by the artificial darks by several percent with this 
method.  

Fortunately, cosmic-ray persistence affected only a small minority of our 
images.  The persistent pixels have charge deposited in them from a cosmic-ray 
hit, which depletes roughly exponentially with time.  Occurrence of 
cosmic-ray persistence is strongly correlated with the {\it HST's} passage through the 
South Atlantic magnetic anomaly (SAA).  
Data for one of our targets, 3CR~65, were severely affected by post-SAA 
persistence, and we have used an experimental method to attempt to 
correct for this effect.  
To do so, we ``sacrificed'' some of the exposures closest in time to the SAA 
passage, and used these as a model of the persistence pattern.  
We then scaled and subtracted this persistence model, minimizing 
the background RMS in a fashion similar to that used by PEDSKY.  
This worked at the $\approx 10\%$ level providing significantly 
improved noise and cosmetic characteristics.  A more robust method 
using darks automatically scheduled after SAA passages 
has been implemented for observations with NICMOS in {\it HST} Cycle 11.  

The individual exposures for each galaxy were combined using the 
DRIZZLE task \citep*{FH02}, with relative shifts determined using image cross-correlation.
Input weight maps (inverse per-pixel variance) were generated with a 
noise model which includes the 
contributions from the sky, readout amplifier glow, standard dark current and read noise.
The DRIZZLE task then generates an output weight map based on the combination
of input error maps, transforming the input in exact analogy with the data themselves.
These output weight maps were used to calculate our error estimate in the 
subsequent analysis.  DRIZZLE does not 
account for the correlation of the noise between adjacent pixels when it 
creates the output weight map. To 
correct the variance for inter-pixel correlation, 
we applied a single multiplicative factor (= 1.84) based on the equation 
derived by \citet*{CdMD+00}.
The effective plate scale after drizzling is 
one-half the original scale, or 0\farcs0375 / pixel 
(the diffraction-limit of {\it HST} at $1.6\mu$m is $\approx 0.168\arcsec$).  
Several original tasks were created for these reductions 
are now included in the STSDAS reduction package.

The WFPC2 frames included in the present analysis are from 
GTO proposals (1070 and 6235; PI M. Longair) from the \HST\ archive 
and several GO proposals (5429, PI W. van Breugel; 5925, PI P. Eisenhardt; 
5967, PI M. Dickinson; 5995, PI O. Le F{\`{e}}vre).
Reduction of the WFPC2 images was done using the standard STSDAS 
pipeline.  The WFPC2 thumbnail images presented in this work have 
been registered with the NICMOS images and interpolated to the 
scale of the drizzled NICMOS images.  Table~\ref{magtab} 
presents magnitudes derived from the reduced data using
aperture (Col.~3-4), total (Col.~5),
and point source (Col.~6) photometry.
These numbers have relatively 
little dependence on the model fits described in the next sections.

\section{Analysis}\label{Analysis}

\subsection{Profile Fitting\label{Fitting}}

To quantitatively compare the morphologies 
of our galaxies to other samples, we fit analytic surface-brightness 
models to the data, primarily concentrating on exponential disk 
and \rq-law \citep*{deV48} profiles.  
The fits were used to determine characteristic sizes and to differentiate 
between the spheroidal and disk models.  
In addition, a Sersic profile \citep*[$r^{1/n}$-law; ][]{SER68} was also fit 
to discern any deviations from the \rq and disk models.  
We decided not to fit more complex models (such as the ``Nuker'' law of
\citealt*{LAB+95}) with more free parameters, because our data are unable to constrain many 
of the added parameters.  For instance, the ``break'' radius in the 
``Nuker'' law is empirically a $100-300$ parsecs, which 
is smaller than the physical resolution of our data.  
We used both one-dimensional (1D) and two-dimensional (2D) fitting algorithms to derive the 
best-fit profile parameters.  
For the 1D fits, the initial source position angle (PA) and ellipticity 
were fixed to measurements using the STSDAS task ELLIPSE.  
The degree of control provided by ELLIPSE allows easy quantification 
of the ellipticity and PA.  
For the 2D method, we performed fits allowing the PA and ellipticity to 
vary and also fixing them to the values determined from our ELLIPSE
measurements.  We subsequently used the fit with the 
lower reduced chi-square ($\chi_{\nu}^{2}$) value.  As it turned out, all galaxies were 
better fit using the fixed ellipticity and PA model.  

The cases of 3CR~470 ($z=1.653$) and 3CR~256 ($z=1.819$) are 
complicated and warrant separate analyses.  Our 3CR~470 
images are 
contaminated by the presence of 
strong H$\alpha$ emission in the bandpass.  
It is also not clear which of the three sources we detect is the radio 
source counterpart.  
3CR~256 is certainly the most 
peculiar, as well as the highest redshift, object in this sample.  
Its peak surface-brightness and total magnitude 
are much fainter than those of our other sample galaxies (see Table~\ref{magtab}).  
3CR~256 is also known to lie well-below the \Kz\ relation formed by other powerful radio 
galaxies \citep*{CE91,SEA+99}.
Although we present the imaging data, we have chosen not to include 3CR~470
nor 3CR~256 in the current analysis.  

\subsubsection{One-dimensional Fits\label{1DFit}}

To fit analytic profiles to the data, we first constructed 
azimuthally-averaged profiles (i.e., one-dimensional 
functions of semi-major axis) for 
both the real data and a set of artificial galaxies.
The analytic galaxy templates span 
a wide range of effective radii (a grid of 200 bins ranging from $10-150$ 
drizzled pixels, or $0.375-5.625$\arcsec) and ellipticity 
(a grid of 5 bins ranging from 0.05 - 0.7).
These
were convolved in two dimensions with a simulated NICMOS Camera 2 PSF.  
We used the TinyTim program \citep*{KRI95}
to generate this PSF on a sub-sampled grid corresponding to our 
final DRIZZLE pixel scale.

\begin{inlinefigure}
\epsscale{0.75}
\plotone{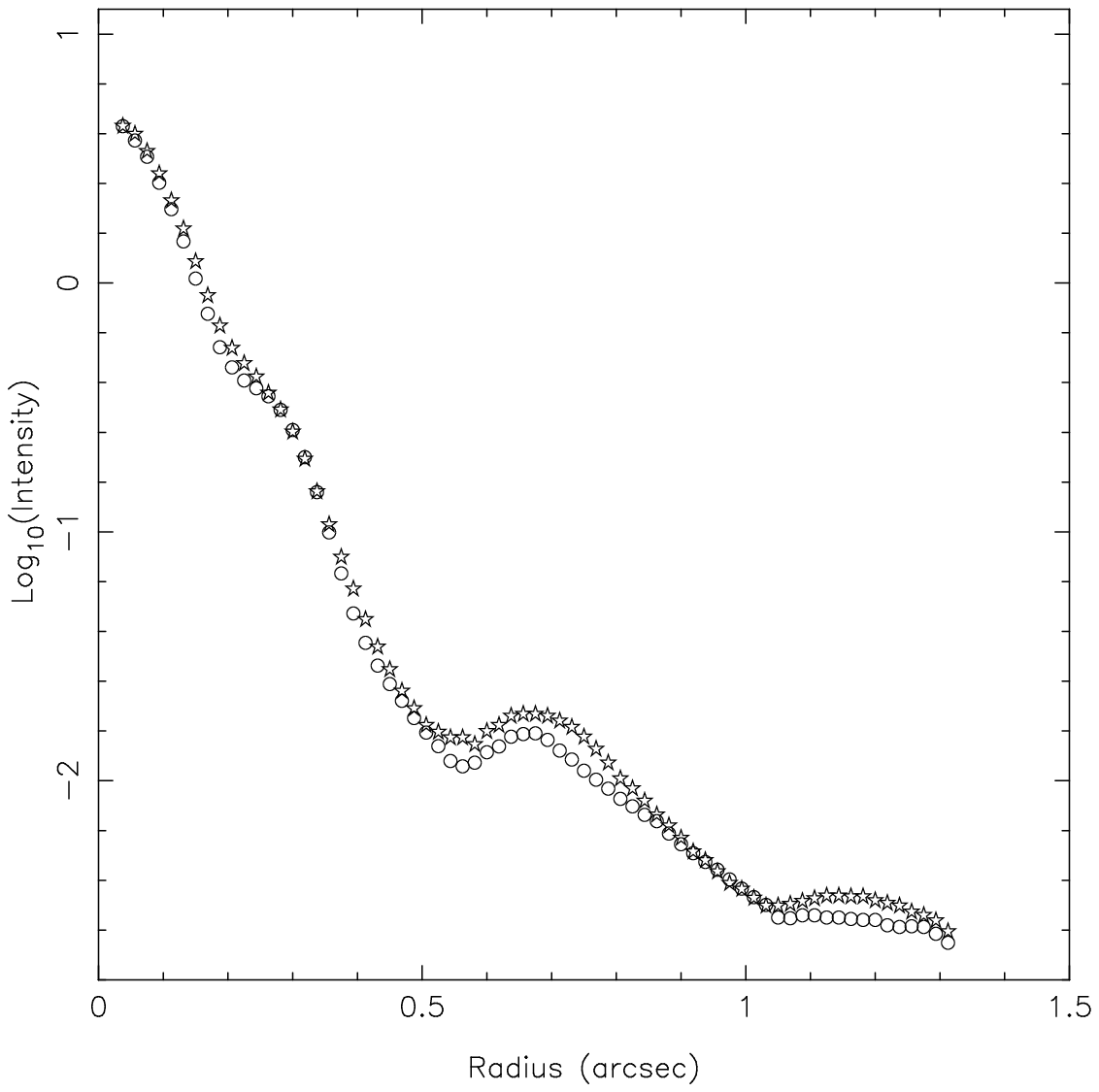}
\figcaption{Azimuthally-averaged intensity versus radius for both the 
TinyTim model PSF (circles) and a bright star in the field of 3CR~470 (stars).  
The two curves match one another well at $r < 0\farcs5$.  The 
discrepancy at larger radii will only affect the photometry of bright 
point sources, i.e., ones much brighter than those found in our sample.\label{psf}}
\end{inlinefigure}

We then smoothed this PSF with a boxcar filter 
representing the original pixel scale to emulate the effects of 
under-sampling due to the finite pixel size.  The resulting PSF has 
approximately the same flux distribution as the final reduced 
data.  Optimally, we would have liked to use an empirical PSF targeted with a similar 
observing mode.  Unfortunately, only a few bright stars exist in our targeted 
fields and we did not propose for separate PSF exposures.  
Comparison with a bright star in one field showed that the TinyTim model
PSF reproduced the stellar PSF well ($\pm 5\%$ in intensity; see Figure~\ref{psf}) out to a radius 
of about 3\arcsec.  
The final PSF-convolved template library consists of 1000 model galaxies from which 
azimuthally-averaged profiles have been extracted.  Then a  
subset of 200 are chosen for each galaxy based on its measured ellipticity.  

We used a $\chi^2$-minimization algorithm 
\citep*[AMOEBA; ][]{NUMREC} to derive the 
sky value, which is an additive normalization, and a multiplicative 
normalization which gives the best fit to the data.  
Each image should be completely sky-subtracted by the 
pedestal removal process (as the small fit sky values confirm); 
varying the sky value only lengthened the fitting process slightly while 
maintaining the greatest possible source-to-source consistency.  
We calculated the input errors for the $\chi^2$ procedure by 
measuring the output weight maps from DRIZZLE 
in the same elliptical isophotes as the galaxy.  
The best-fit model for each source was determined by comparing 
the resultant $\chi^2$ values for each of the 200 templates.
The central $\approx$0\farcs5 diameter has been excluded from these fits to 
reduce the influence of any nuclear point sources and uncertainties 
associated with the PSF and centering.  
\vspace{3.0 truecm}

\subsubsection{Two-Dimensional Fits\label{2DFit}}

Given a set of input parameters,  
the two-dimensional fitting algorithm 
generates model images ``on-the-fly''.
The freedom of the 2D approach allows us to fit for any 
combination of galaxy centroid, position angle, 
ellipticity, effective radius, surface-brightness, point-source
contribution, and sky value.  
The method can be easily extended to any analytic profile.
We primarily attempted to distinguish between 
de Vaucouleurs and exponential 
disk profiles, but also fit a $r^{1/n}$ Sersic law to determine whether 
deviations from the \rq or disk models were prominent.  
We did not fit more complex profiles 
with additional free parameters because we 
felt it was not warranted by the signal-to-noise ratio 
and physical resolution of our data. 

The program extracts a thumbnail image centered 
on the galaxy from the larger, sky-subtracted mosaic.  
Initially, this sub-image was 200 drizzled pixels 
($7\farcs5$) on a side, with the outer fit radius 
fixed to half this value ($3\farcs75$).  Trials with an outer fit 
radius twice as large ($7\farcs5$) did change the 
fits slightly for several galaxies.
The average decrease in effective radius, $R_e$, was $\approx$ 10\%. 
We chose one of these outer fit radii based on a simple by-eye comparison 
of the residuals for each source, rather than apply 
a strict test of $\chi_{\nu}^{2}$ values.

The algorithm then refines the galaxy centroid to sub-pixel precision.
An initial guess at the fit parameters is passed to a function that generates 
a model image which is then convolved with the TinyTim PSF (described above).
The model is calculated on a sub-sampled pixel grid.  Our 
sub-sampling by a factor of five allows the surface-brightness 
per pixel to be calculated to higher precision than at the original scale.
This also allows centering to be accurate to less than a pixel.  
The model is then rebinned 
to the size of the input data thumbnail while conserving intensity.

A $\chi^2_\nu$ is calculated for 
the model fit using the output weight map derived from the DRIZZLE task 
and a binary mask generated to remove neighboring objects and to impose 
inner and outer fit radii.
We masked neighboring objects or high surface-brightness aligned light 
in the following galaxies: 3CR~65 (neighbor), 3CR~184 (neighbors), 
3CR~210 (neighbors), 3CR~239 (small neighbors), 3CR~266 (neighbors and aligned 
light), 3CR~267 (neighbors), 
3CR~280 (neighbors and aligned light), and 3CR~368 (cores of neighboring, largely 
subtracted stars and aligned light).  
The Levenberg-Marquardt minimization algorithm \citep*{BEVINGTON} is 
used to generate the next guess for model parameters until the fractional difference between 
successive $\chi^2_\nu$ values is less than a small user-defined value.

After testing several combinations of fit parameters (see \S\ref{simulations}) 
we also included a point source component 
in our fits and reduced the inner fit radius to zero.
The fact that there is a degeneracy between the central cusp of a 
de Vaucouleurs profile and a nuclear point source, is
discussed in more detail in \S\ref{PtSrc}.  
The simulations we performed convinced us of the importance
of including this component.  
The galaxy centroid was allowed to move for all of the 2D fits.  
The point source centroid was always fixed to the galaxy centroid.
In addition to analyzing our own data,
we have used the two-dimensional method to 
fit 15 intermediate redshift radio galaxies from 
the 3CR WFPC2 snapshot survey \citep*{dKBS+96,MBS+99}.  
The results of the 2D fits are presented for the high- and 
intermediate-redshift RGs in Tables~\ref{partab} and \ref{snaptab} 
and are discussed in \S\ref{ResDisc}.  
\begin{figure*}[t]
\hspace{-7.5in}
\plotfiddle{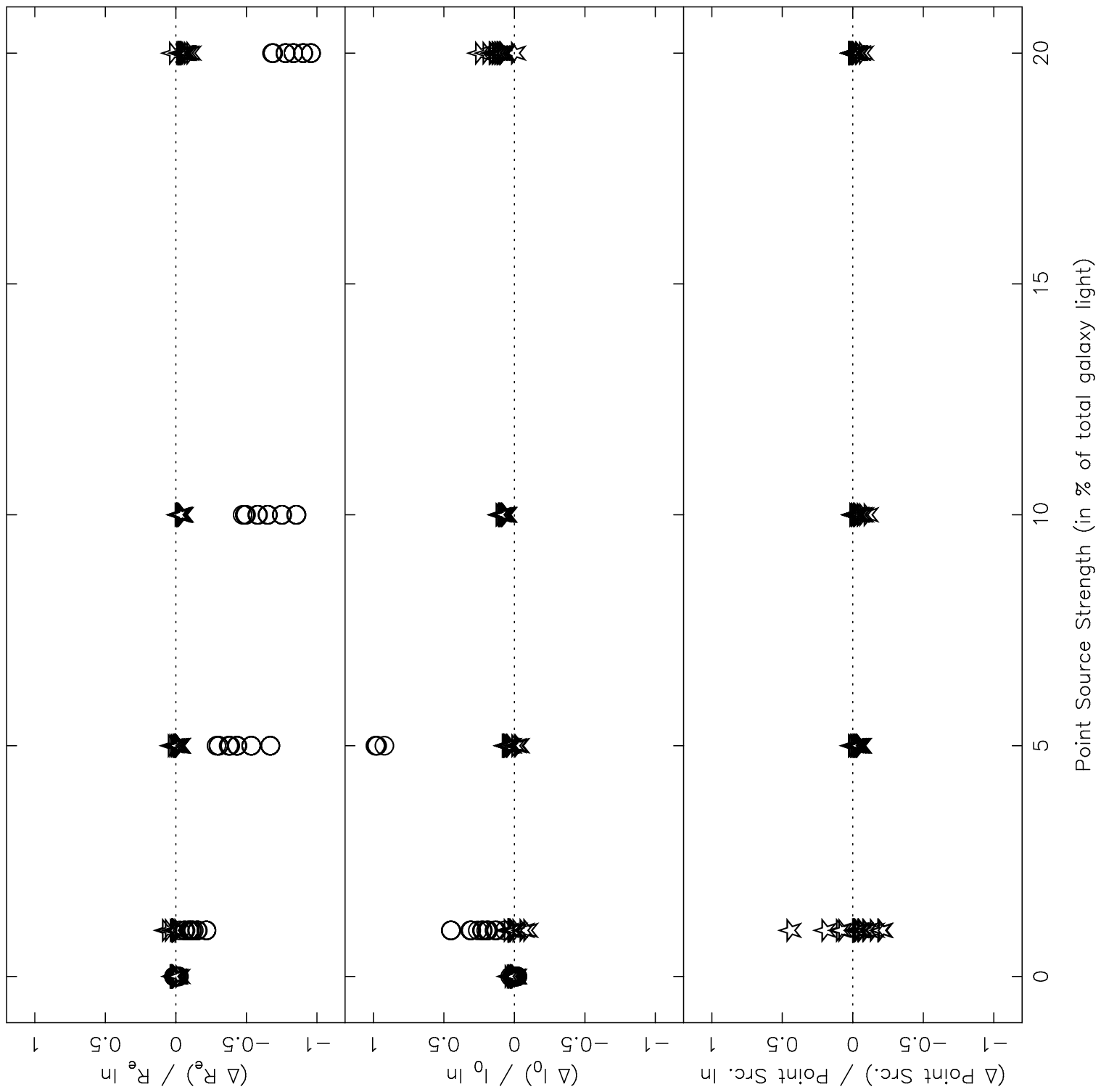}{0in}{-90.0}{0.75}{0.75}{}{}
\figcaption{Percent deviations (output minus input) 
of derived parameters from input parameters as a function 
of input point source luminosity.  The stars indicate fits which include a point 
source contribution in the models, the circles have only an \rq-law component.  These 
values were derived using the simulations fully described in \S~\ref{simulations}
\label{simratiofig}}
\end{figure*}

\subsection{Simulations and Error Analysis\label{simulations}}

To better understand the systematic uncertainties in the 
fits, we undertook a thorough set of simulations.
The full parameter space was explored by 
creating a large library of artificial galaxy images
distinct from those used in the 1D fits.
Models were generated for a wide range of effective radius, nuclear point-source 
luminosity and signal-to-noise ratio, which is defined per pixel at $R_e$.  
We assumed a constant sky background, but the noise was modulated by the 
final weight-map produced by DRIZZLE (from a representative 
sub-image of a single target's weight-map).  
These galaxies were then fit using the 2D method described above.
Position angle and ellipticity were held constant at the same input value 
for all simulated sources.  Table~\ref{simtab} shows the range of input 
values for the model galaxies.

In Figure~\ref{simratiofig} we show
the fractional deviation of the output fit values 
from the input values for radius, surface-brightness, and point-source strength 
versus point-source luminosity, which is shown as a percentage of the total 
galaxy luminosity.
We can estimate the systematic errors 
on all the fit parameters by measuring 
the deviations of the output values compared to the input.
Systematic errors on the primary fit parameters, effective radius, $R_e$, and 
effective surface brightness, $I_e$, are less than
5\% for all model galaxies with signal-to-noise ratio $\geq$ 1 
measured per pixel at $R_e$.  
The median signal-to-noise ratio for our sources is $\approx 3.5$. 
Therefore, we expect the systematics on 
our derived parameters to be roughly in line with those simulated here.  
In Figure~\ref{simratiofig} stars represent fits 
which include a point-source component, and 
circles represent fits which are purely de Vaucouleurs law models.  
The inclusion of a point-source 
in the models greatly reduces the systematic error on all three derived parameters:
$R_e$, $I_e$, and, of course, point-source luminosity.  
For point-source luminosities less than $\approx 10\%$ of the total galaxy 
light, 
the point-source flux itself is determined to $\approx 5$\% of its input value.
This is comparable to the intrinsic scatter for the galaxies with only a small 
point-source contribution.  Therefore, the error in the nuclear luminosity determination 
may introduce a spurious contribution to the total error 
budget when the nucleus is not very bright.  

\begin{figure*}[t]
\epsscale{1.0}
\plotone{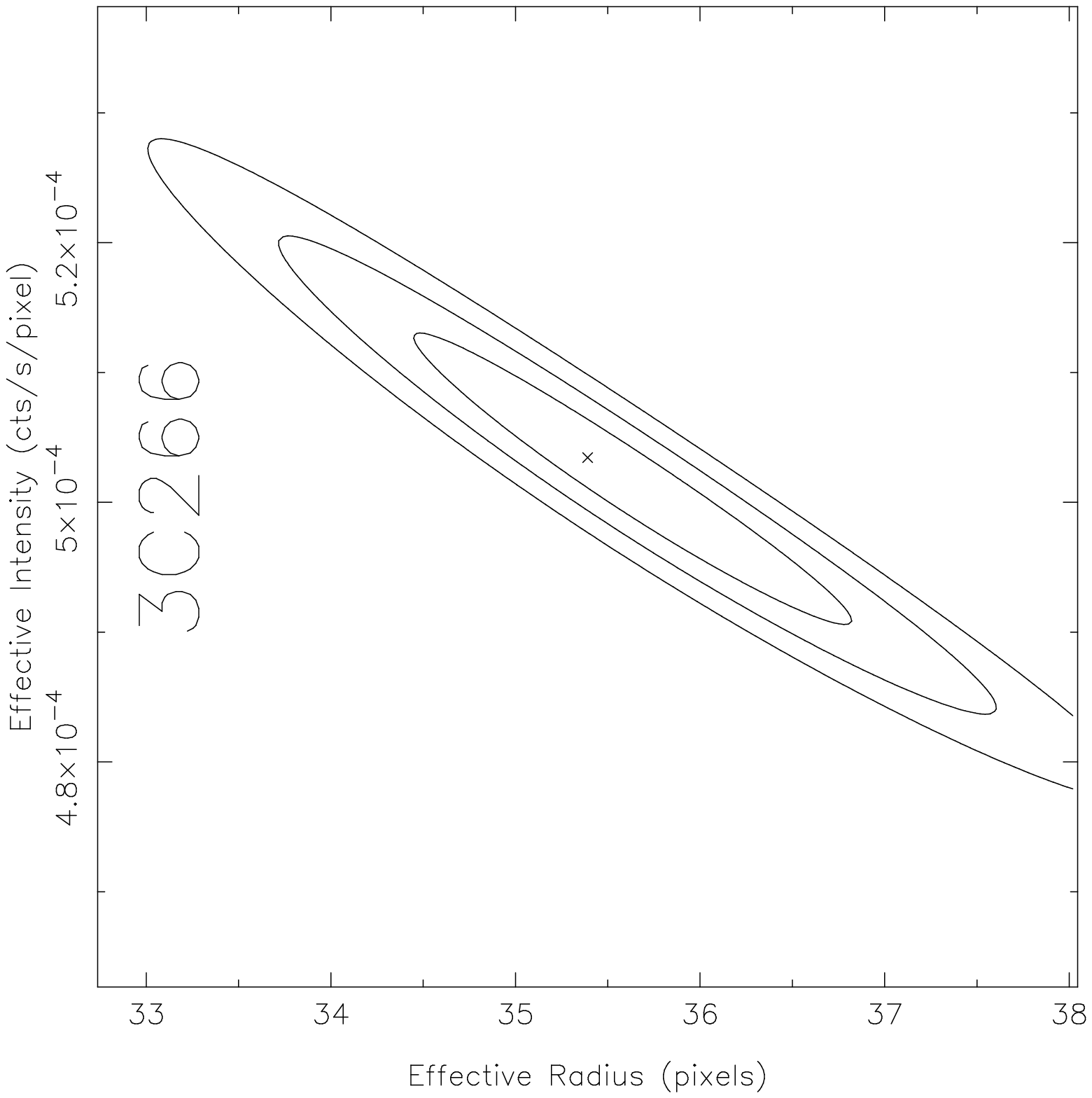}
\caption{$\chi^2$ contours representing one, two and three $\sigma$ errors.  
The axes are linear intensity and radius, 
and the box is 15\% of the best-fit value 
on a side.  The best-fit value is marked with a cross.  
All of the fit galaxies share a similar $\chi^2$ and error 
contour structure.  The 1$\sigma$ error is $\approx 8\%$ 
of the best-fit value in intensity and $\approx 5\%$ 
of the best-fit value in radius.
\label{errorcontourfig}}
\end{figure*}

Based on these simulations, we expect to 
underestimate the point-source contribution, and therefore 
underestimate the effective radius, because a more concentrated profile 
accounts for the ``missed'' point-source light. 
The simulations can be used to quantify this effect 
as a function of point-source luminosity.
For the derived point-source strengths of our targets, this is not a significant 
effect and no correction has been made.
Therefore, we have chosen to perform fits to the data both with and 
without a point-source component, and determine which produces the best 
residuals, i.e., the lowest $\chi^{2}_{\nu}$.  

We measured the random errors using a $\chi^2$ technique.  For each target galaxy 
we calculated $\chi^2$ values for a fine linear grid of $R_e$ and $I_e$ 
centered on the best fit values.  
All objects share essentially the same $\chi^2$ structure.  
A representative contour plot is shown in 
Figure \ref{errorcontourfig}.  The box is $15\%$ of the best fit values on each side, with 
radius the abscissa and linear intensity the ordinate.  The contours represent 
the gaussian one, two, and three $\sigma$ errors based on $\Delta \chi^{2}$ values of 
2.3, 6.17, and 11.8 respectively as calculated for the standard two 
degree-of-freedom $\Delta\chi^{2}$ table \citep{NUMREC}.  The median 1$\sigma$
error for all the sources is roughly 10\% of the 
best-fit value in both radius and surface-brightness.  

The one- and two-dimensional fits agree reasonably 
well as illustrated 
in Figures~\ref{1d2d_rad} and \ref{1d2d_sb}.  
A linear regression analysis calculates the scatter about the one-to-one 
relation to be $0\farcs15$ in radius and 0.3 mag in surface-brightness.
Recall that these are completely independent methods of determining the fit 
parameters.  
The agreement between the results of the two algorithms in addition to the results of the 
simulations gives us confidence that large, systematic 
errors in our derived galaxy parameters are unlikely.  

\begin{inlinefigure}
\epsscale{0.6}
\plotone{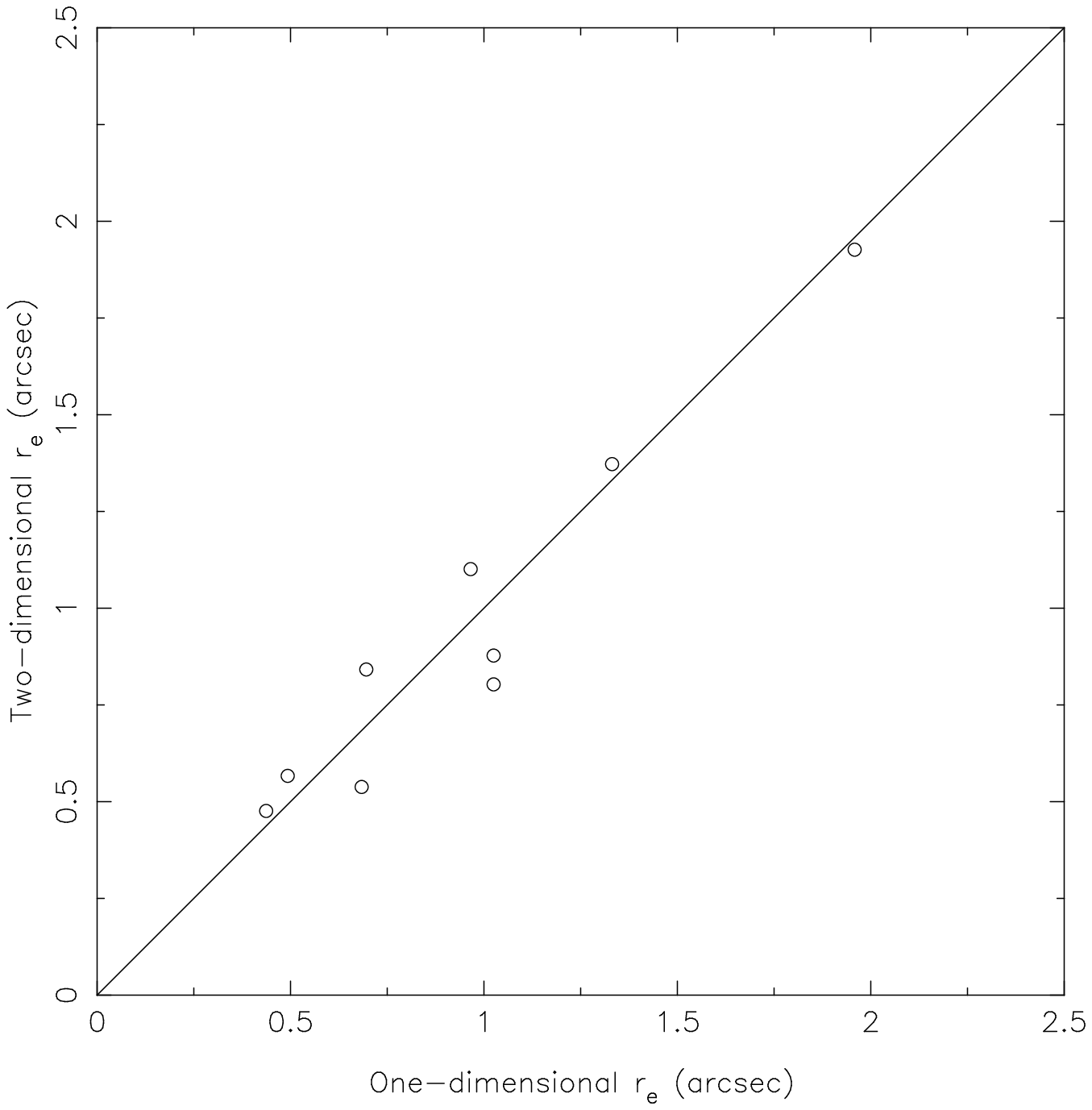}
\figcaption{Effective radius derived via the 2D fitting method versus radius 
derived via the 1D algorithm.  The solid line is the one-to-one line.
The two independent methods show good agreement with a scatter about the 
one-to-one line of $0\farcs15$.
\label{1d2d_rad}}
\end{inlinefigure}

\begin{inlinefigure}
\epsscale{0.6}
\plotone{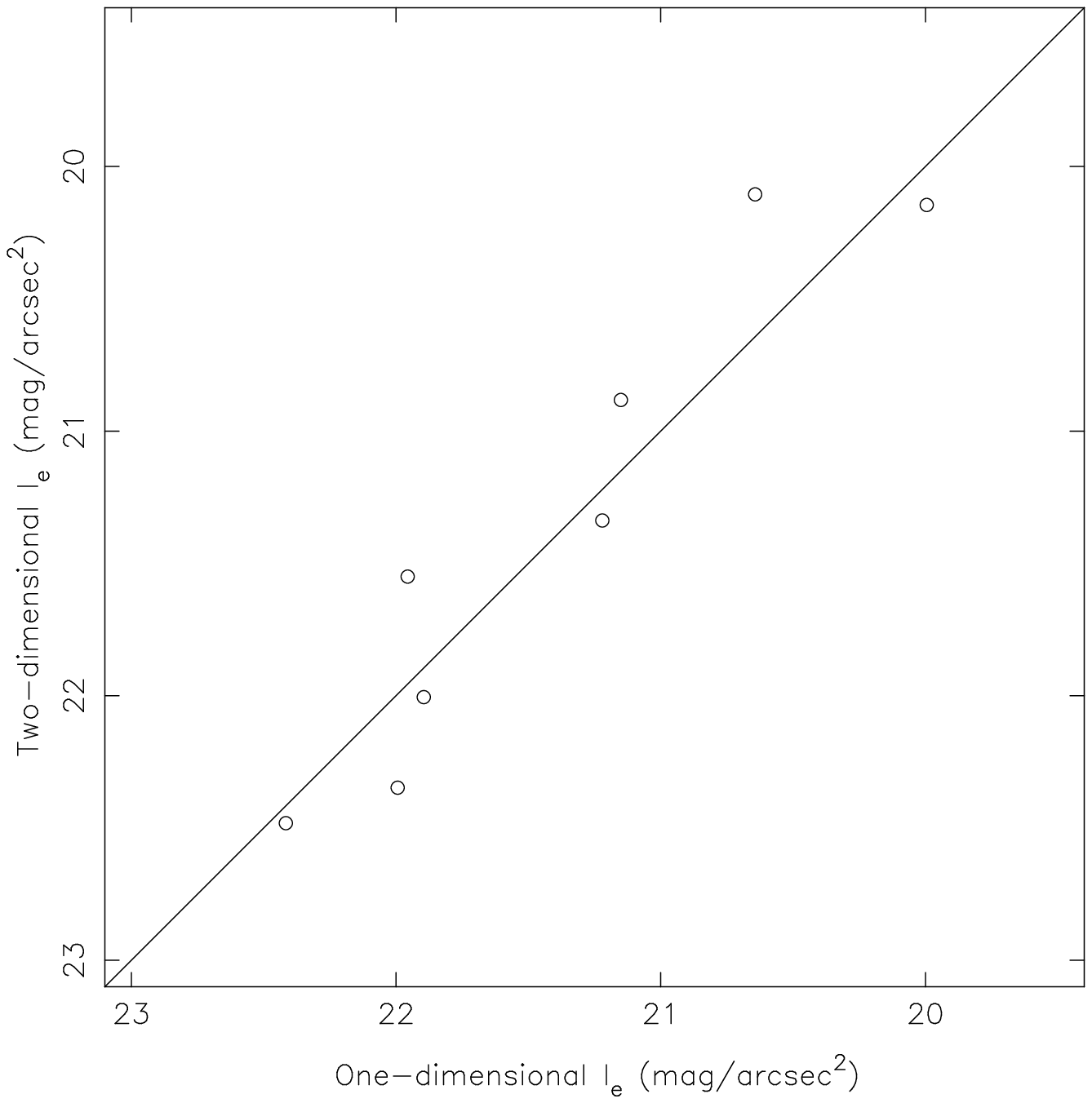}
\figcaption{Effective surface brightness from the 2D fits versus the same from the 1D fits.
Again, the two independent fitting methods show good agreement.  The scatter 
about the one-to-one line is 0.3 magnitudes.
\label{1d2d_sb}}
\end{inlinefigure}

\section{Results and Discussion}\label{ResDisc}

\begin{figure*}[t]
\hspace{-7.5in}
\plotfiddle{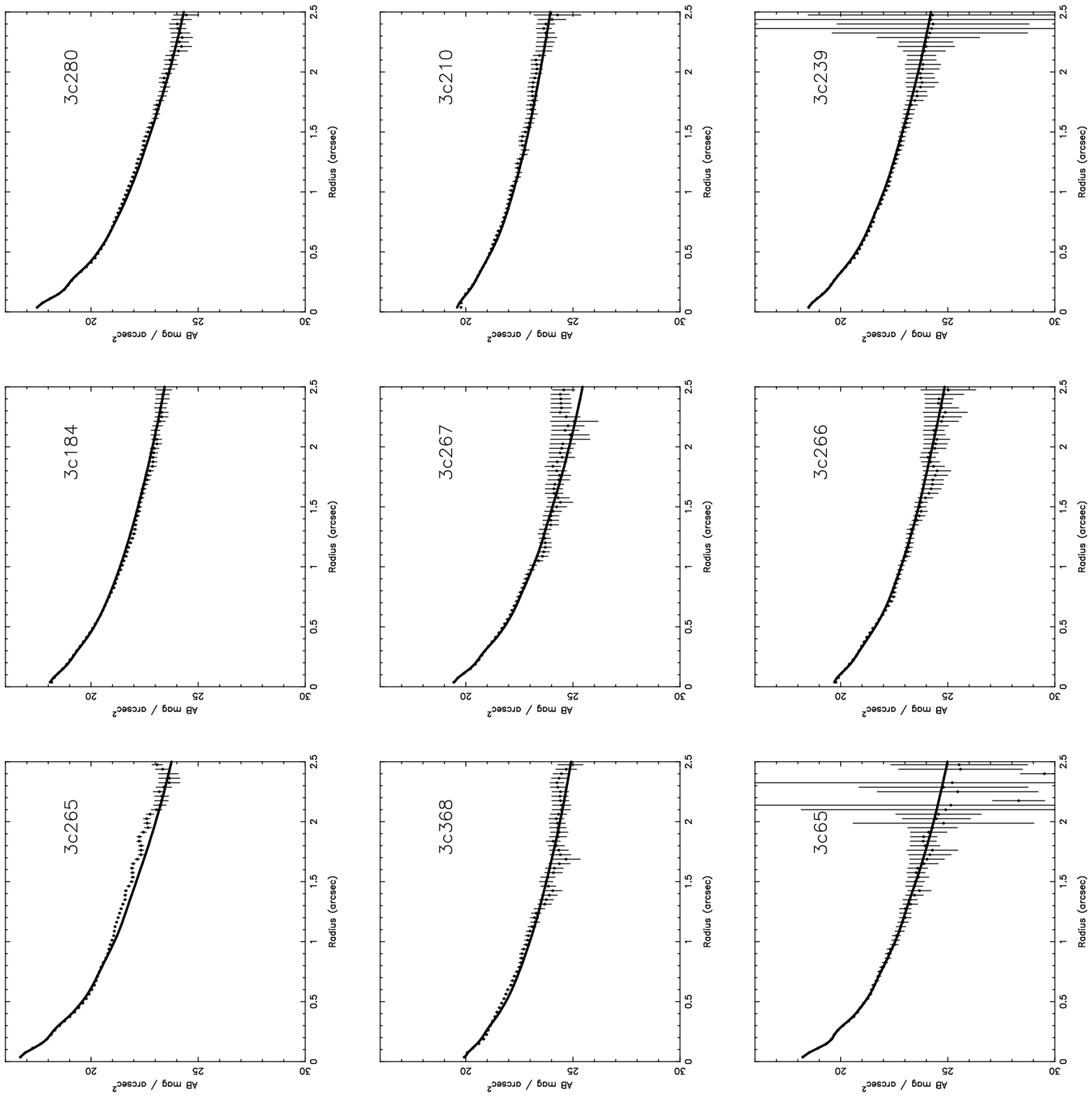}{0in}{-90.0}{0.8}{0.8}{}{}
\figcaption{Surface-brightness profiles with derived two-dimensional fits.  
The solid line is the best two-dimensional \rq fit.
The high central surface-brightness, diffraction rings and inclusion of a 
point-source in their best-fit models imply that 3CR 65, 265 
and 280 may harbor obscured nuclear sources.  
\label{ProfileFig}}
\end{figure*}

The NICMOS data presented here are unique in their combination 
of high spatial resolution and NIR wavelength coverage.  
Previous studies of $z \approx 1$ radio galaxies were limited either 
by contamination from the AGN at observed optical wavelengths,
i.e., the UV alignment effect, or by ground-based seeing in the NIR.  
The exquisite resolution and NIR sensitivity provided 
by \HST\ has allowed us to analyze these data in ways that were 
previously impossible or unwarranted.  In this section we discuss the 
results of our analysis, focusing first on the galaxy morphologies and 
then on their comparison to other galaxy samples.  The key 
quantities used in the following discussion are presented in Tables~\ref{magtab} and 
\ref{partab} (and Table~\ref{snaptab} for comparison).  
Table~\ref{magtab} contains various magnitudes which 
are either model-independent or depend only weakly on the 2D fits, while Table~\ref{partab} 
presents parameters derived directly from the fits which may depend 
on our choice of cosmological model.

\subsection{Host Galaxy Morphology}

\begin{figure*}[t]
\epsscale{1.0}
\plotone{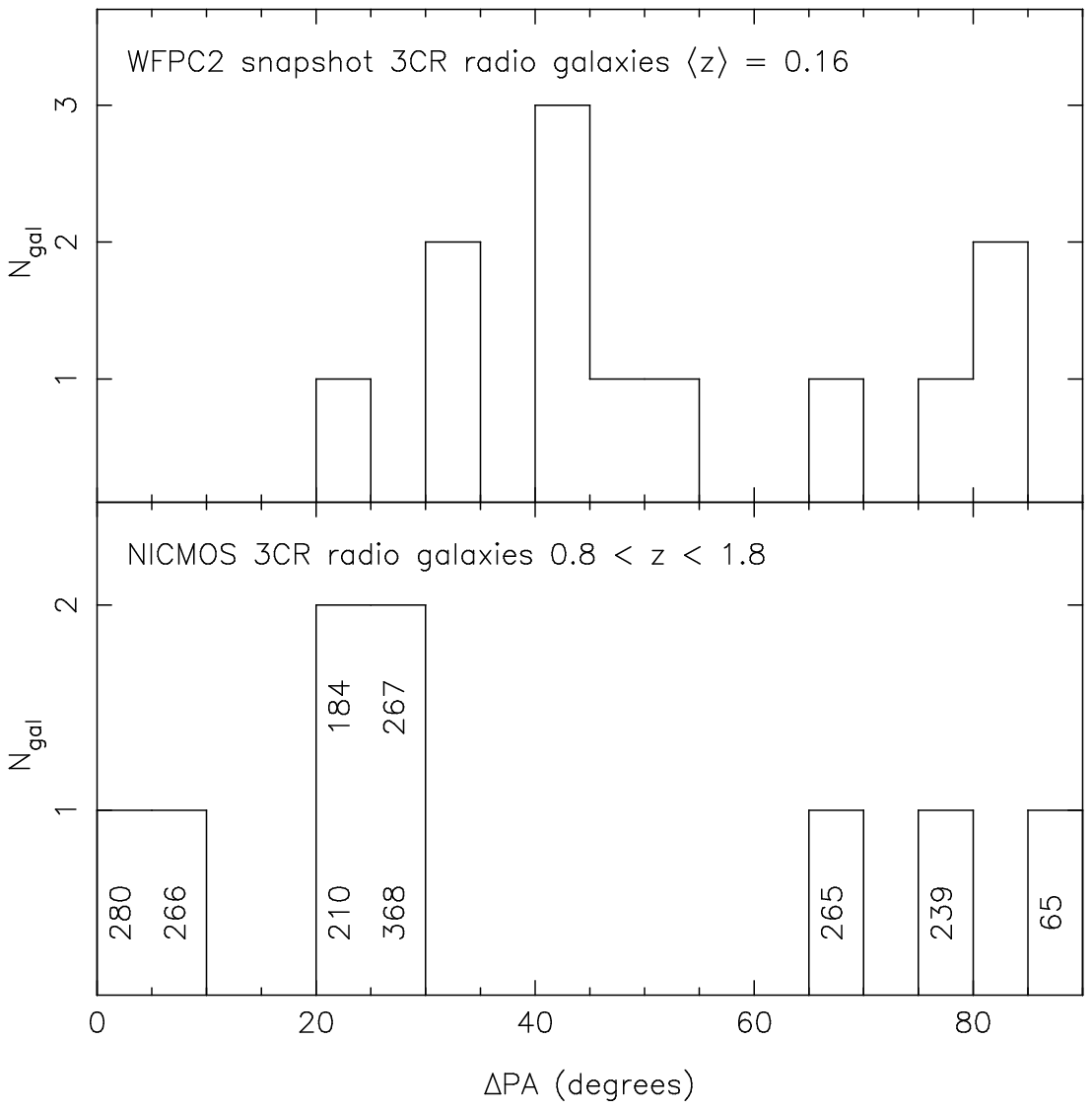}
\caption{Difference between the position angles of the radio source 
and the host galaxy major axis, measured at optical rest-frame wavelengths.  
The top panel shows measurements from WFPC2 images of low-redshift 
radio galaxies, while the bottom shows NICMOS results for our 
high-redshift sample. The lower histogram is labeled with the target 
names.\label{pa}}
\end{figure*}

The NICMOS images (Figure~\ref{TrioFig}) reveal 
round, symmetric host galaxies unseen in the UV WFPC2 data.  
Given the high angular-resolution of NICMOS, this unambiguously 
confirms the results of previous authors \citep[e.g.,][]{RLS+92,DDS95,BLR98}.  
The signal-to-noise ratio and spatial resolution of the data allow a 
detailed investigation of the varied luminous components.  
In this section we present the analysis of the host galaxy 
morphologies.  We studied the extracted 
profiles in detail to look for signs of both nuclear point 
sources and non-homology (i.e., the degree to which the profiles differ from 
one another).  In combination with the comparably resolved 
UV data from WFPC2, the correspondence between 
the AGN-related aligned light and the stellar host 
can also be determined.

\subsubsection{Stellar Component}

The one- and two-dimensional surface-brightness profile 
fits statistically prefer the \rq models over the exponential 
disk profiles.  
To test this preference we 
have employed the statistical F-test to compare the resulting 
$\chi_{\nu}^2$ values.   
The disk model can be rejected 
in favor of the de Vaucouleurs profile 
at greater than $99.9\%$ confidence for \emph{all} objects.
This suggests that the host galaxies are indeed ellipticals
and provides circumstantial evidence that they 
are the passively evolving galaxies 
responsible for the smooth \Kz\ relation.  
The NICMOS data presented here are the most 
direct detection of the hypothesized bulge 
component.  The correlation between bulge and black-hole mass 
\citep{MTR+98,GBB+00,FM00} also predicts 
that very luminous AGN should reside at the centers 
of large, bulge-dominated galaxies.  
We note, however, that the de Vaucouleurs model is an 
oversimplification which ignores several 
details of the surface-brightness profiles.

Figure~\ref{ProfileFig} shows the smooth, simple surface-brightness profiles 
of the nine fitted galaxies.  The solid line is the 2D \rq fit.  
Close inspection of the profiles reveals several features.  
Three galaxies, 3CR~265, 3CR~280 and 3CR~65, show very steep central cusps, high central 
surface-brightness and the hint of the first diffraction ring of the 
NICMOS Camera 2 PSF (at $r \approx 0\farcs25$).  
These clues suggest the presence of nuclear point sources
or at least a very cuspy profile. This possibility we 
investigate more fully below and is included in the 2D fit 
for these 3 galaxies.
3CR~368 and 3CR~210 show very flat outer profiles, seemingly 
more consistent with a power-law model rather than a de Vaucouleurs profile.  
Also, in both 3CR~265 and 3CR~280 the data exceed the \rq profile at 
radii greater than $\approx 0\farcs7$.  This may be due to their 
extremely high central surface-brightness, because
to compensate for the central surface-brightness, 
the \rq law must be more compact and 
therefore underpredicts the outer intensity.  
However, this excess light at large radii may 
also be indicative of extended, cD-type halos (both 265 and 280 are believed to 
lie at the centers of moderately rich clusters), extended emission-line nebulae 
(for example, [SIII]~$\lambda 9533$ in the case of 265) or the 
aligned light (particularly the ``arc'' to the 
west of 280, which was masked in the 2D fits but may still 
contribute measurable diffuse flux).  

Perhaps the more significant conclusion to be drawn from the 
differences between the galaxy profiles is that these 
sources may not form a homologous sample.  
Homology is relevant to the various scaling relations 
followed by early-type galaxies.  For instance, the 
Fundamental Plane is tilted with respect to the 
predicted ``virial plane'' \citep*[e.g.,][]{BCC+97}. One explanation for the 
tilt is non-homology among the galaxies.  
We have fit a Sersic profile to each source in order to 
test the assumption of homology for our sample. The Sersic profile 
\citep{SER68} is defined:
\begin{equation}
I(r) = I_e  e^{-b_n \times[ (r/r_e)^{1/n} - 1]}
\end{equation}
where $b_n$ is calculated such that $r_e$ is the half-light 
radius. This more general profile has been shown 
to correlate with physical galaxy properties 
such as stellar velocity dispersion (see \citealt*{GEC+01,TGC01}).  
Sersic fits performed in the 
same manner as the \rq and disk fits reveal a range 
of the index $n$ within our sample: 
from 2.6 for 3CR~266 to 6.8 for 3CR~280 (see Table~\ref{partab}).  
Due to the limited radial extent of the data, these results 
should be considered suggestive rather than conclusive, since the 
best-fit $n$ value depends strongly on the behavior of the profile 
at large radii (particularly for $n > 4$).  Deeper data 
will be able to distinguish between different $n$ with 
greater confidence.  
\citet{GLC+96} have shown that brightest 
cluster galaxies often obey the flatter, $n > 4$ profile.  
It is interesting that several of our targets are 
also known to reside in moderately dense environments, 
including 3CR~265 and 3CR~280, two of the galaxies exhibiting an $n > 4$ 
profile. Again, based on the statistical F-test, 
the Sersic fits do not significantly improve upon the \rq fits 
except in the case of 3CR~184 where $n = 2.6$.  
The range of $n$ within the sample may also partly explain the 
observed change in derived effective radius when the outer 
fit radius is increased.  There is a correlation between 
the derived Sersic index and the degree to which 
the change in outer fit radius affects the measured galaxy size.
Table~\ref{partab} (Col.~8) lists the Sersic indices for 
the galaxies.  

\vspace{2.0truecm}
\subsubsection{Host Galaxy Alignment\label{Alignment}}

\begin{figure*}[t]
\epsscale{1.0}
\plotone{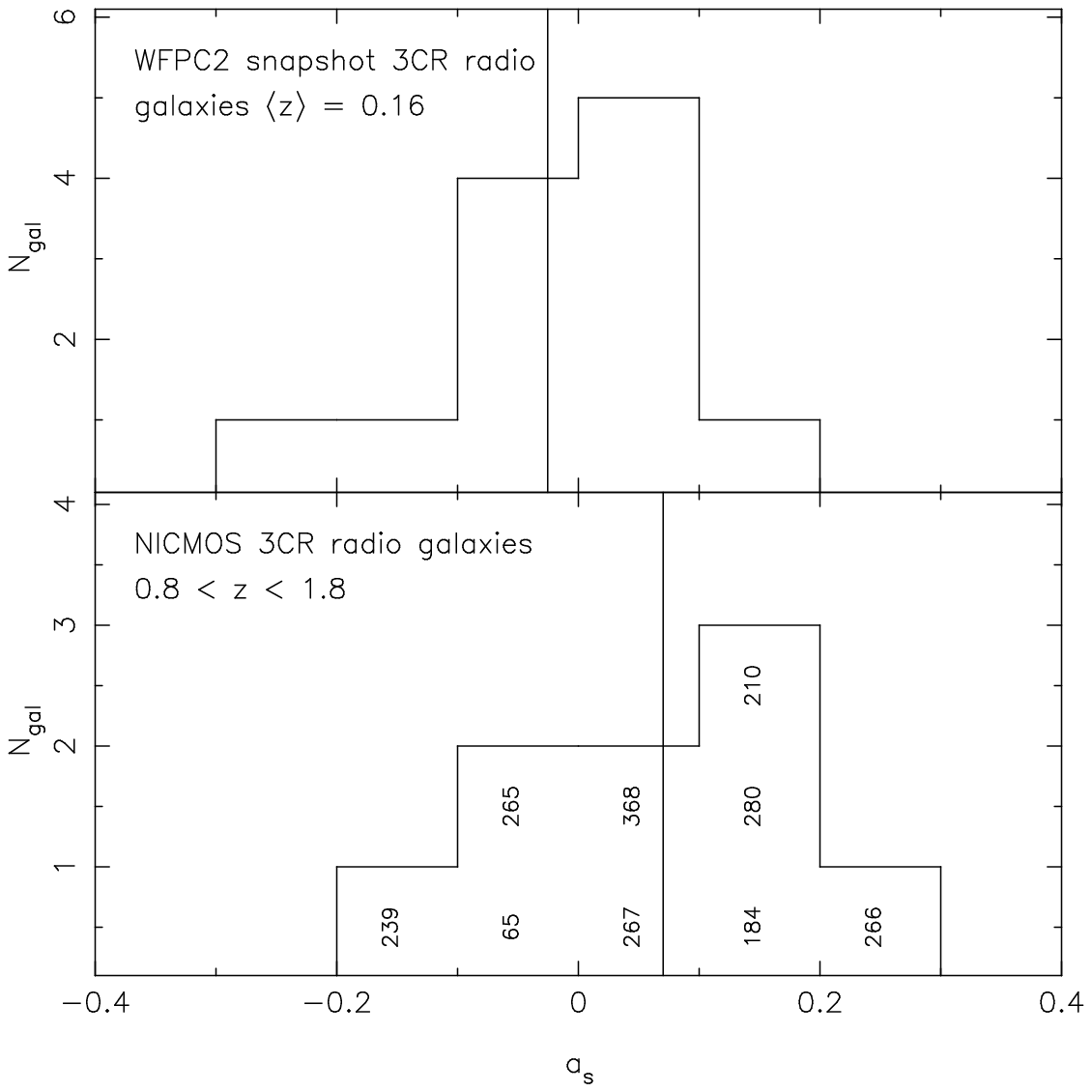}
\caption{Alignment strength (as defined by \citealt{BLR98}) for the $z \approx 1$ 
and intermediate redshift radio   
host galaxies.  Positive $a_s$ indicates an alignment with the radio axis.  The 
high-redshift distribution in the lower panel has both positive mean 
(vertical solid line) and median, 
indicating a tendency for the 
stellar host to align with the radio axis. 
The histogram is labeled with the target names for clarity.   
The lower redshift sample, 
in the upper panel, shows 
no such alignment, with a slightly negative mean (solid line) and median.
\label{as}}
\end{figure*}

Apart from the stellar host, the NICMOS data 
also contain other luminous components, some of which 
are directly comparable to structures seen in the rest-frame 
UV images from WFPC2.  
The residuals shown in Figure~\ref{TrioFig} are faint and 
often aligned with the radio axis, shown as
lines in the middle panels of Figure~\ref{TrioFig}.  
The direct correspondence between the NIR aligned residuals and 
the structures seen in the optical WFPC2 images suggests 
that much of the NIR alignment is simply the long-wavelength 
extension of the rest-frame UV alignment effect.  
The detailed photometric properties of the aligned light will be discussed 
in a future paper.

Perhaps more interesting, however, is the observed alignment 
between the stellar host and the radio axis.
The luminosity-weighted position angle (PA) of the stellar host has been measured 
using the IRAF task ELLIPSE (see \S~\ref{Fitting}) with 
any bright aligned light excluded from the ELLIPSE fitting 
(as described in \S~\ref{1DFit}).  
The host galaxies are certainly less elongated than 
the UV continuum light, but surprisingly many remain within $\sim 30$ degrees of the 
radio axis (Figure~\ref{pa}).  A KS test against a uniform 
distribution shows that the $\Delta$PA distribution is inconsistent
with a random distribution at the 99.9\% confidence level.  
Of course, these measurements of PA depend on the ellipticity 
of the source.  Very round galaxies, such as 3CR~65 and 3CR~239, 
do not provide a meaningful PA.  \citet*{BLR98} defined a metric, 
which they call ``alignment strength'', 
$a_s = \epsilon \times (1 - \frac{\Delta\textrm{PA}}{45})$.  
This properly weights each measurement according to 
galaxy ellipticity ($\epsilon$) and $\Delta$PA.  
A positive value of $a_s$ indicates alignment.  
The $a_s$ values for our sample are plotted as a histogram in 
Figure~\ref{as}.  The distribution has both positive mean equal to 0.07 and 
median equal to 0.06, indicating a tendency toward alignment.  Two-thirds 
of the galaxies show positive alignment strength, similar to the value of 
$70\%$ found by \citet{BLR98}.  

In order to determine whether or not the alignment is 
an intrinsic property of the stellar component 
or the result of contamination by the aligned component 
we performed simulations of randomly oriented galaxies with comparable 
aligned light. The aligned light component is treated using both a model 
and using the actual measured NICMOS residuals for 3CR~266.
The results of these simulations 
suggest that both the ellipticity and PA measurements can 
be skewed by a modest amount of aligned light.  
Therefore, we believe that while these results are intriguing, 
they could reflect some incomplete subtraction of the aligned light 
rather than a property of the stellar distribution.  
However, the same residual aligned light does not have much 
of an effect on the other results presented in this paper, contributing 
on average only 4\% of the total light from the host galaxy.

In addition, using our same software, 
we have measured ellipticities and derived sizes 
and luminosities
for a sample of intermediate redshift radio 
galaxies from the 3CR WFPC2 Snapshot Survey \citep{dKBS+96,MBS+99}.  
The ``snapshot'' sources were chosen 
using several criteria: multiple exposures to allow cosmic-ray removal, extended galactic 
appearance (i.e., not quasars or ``N'' galaxies), 
no major dust features, and good signal-to-noise ratio (which 
eliminates many of the higher redshift sources).  This selection resulted in 
15 3CR galaxies being chosen for fitting: 3CR 219, 234, 310, 314.1, 
319, 332, 346, 35, 381, 403, 436, 452, 63, 79, 93.1.  The median 
redshift of the snapshot sample is 0.16 and the range is 
$0.03 \leq z \leq 0.26$.  All the snapshot galaxies were 
observed with either the F675W or F702W.  
With a slightly negative mean $a_s$,    
these sources do not show the same host galaxy alignment as the 
higher redshift 3CRs.
We revisit these sources for more detailed comparison 
in Section~\ref{Comparison}.  

\vspace{1.0truecm}
\subsubsection{Dust}

\begin{figure*}[t]
\plotone{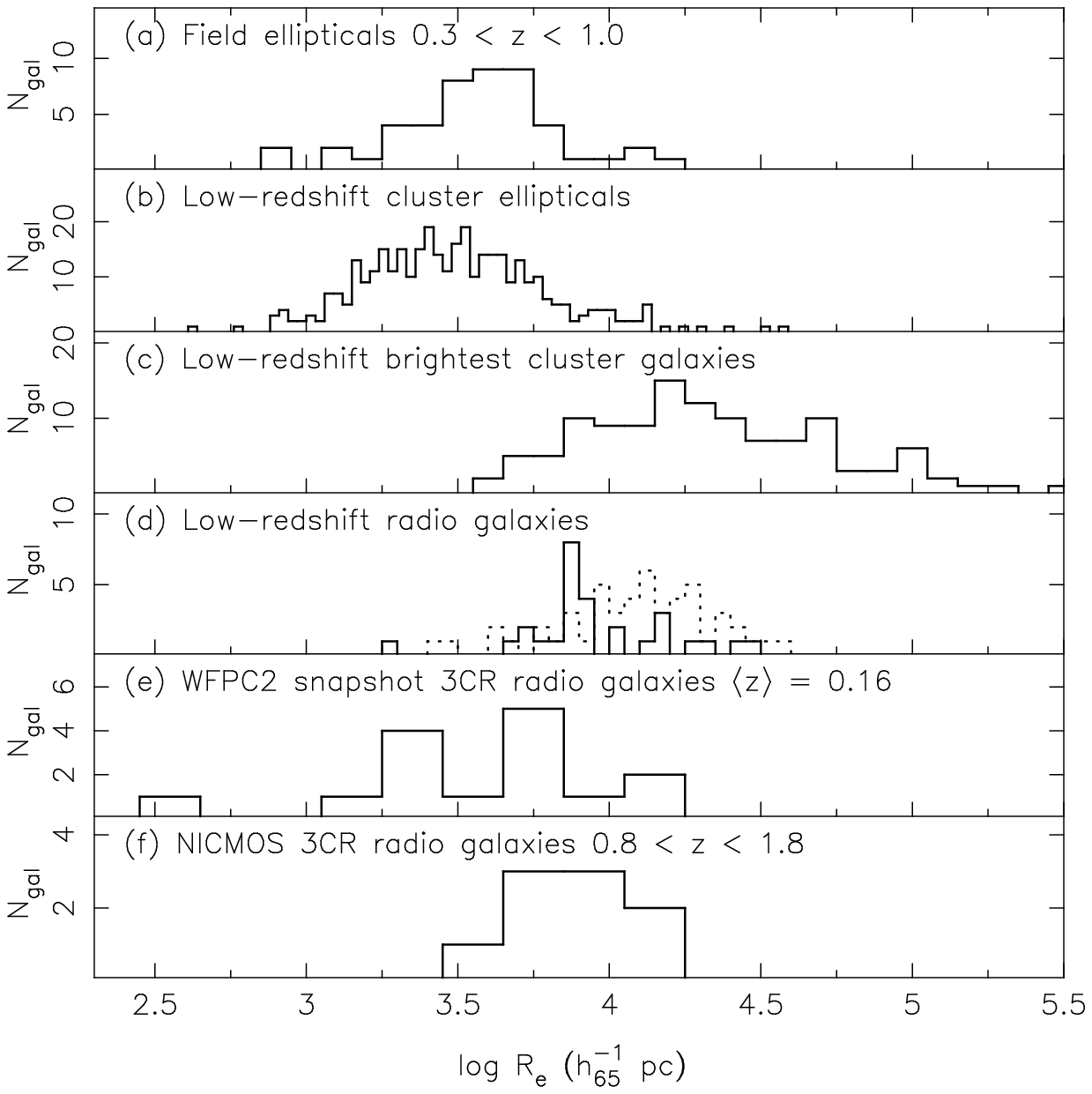}
\caption{Size distributions for the samples described in the text; (a) \citet{SLC+99}, 
(b) \citet{JFK95}, (c) \citet{GLC+96}, (d) \citet{GFF+00} 
(FRIIs, solid histogram, FRIs, dotted histogram) 
and (e) and (f) the intermediate- and high-redshift 3CR galaxies from this work.  
The sizes of our sample galaxies seem to lie most closely 
to the low-redshift FRIIs in panel (d), however, the KS test statistically rejects this 
hypothesis.
\label{sizehist}}
\end{figure*}

The symmetric host galaxy is often centered on a local 
minimum of the UV light (e.g., 3CR~210, 3CR~368), suggesting 
that dust is playing an important role in the 
observed morphology of these sources \citep*[cf.][]{DDS95}.  
In low-redshift radio galaxies, dust is quite common 
in the central region, often configured in either a disk 
or a dust lane \citep[e.g., ][]{MBS+99,dKBB+00}.  In these 
local objects, the dust is known to be associated with the 
central engine.  
The physical resolution of our data is insufficient to 
make the same claim for our targets.  
However, the presence of molecular gas in some intermediate- to high-redshift 
radio galaxies \citep*{PRvdW+00} also implies that conditions suitable for 
dust exist.  
The existence of dust in these sources may also be 
inferred by the presence of scattered light, although electrons 
could instead be responsible.

\subsection{Nuclear Point Sources\label{PtSrc}}

AGN unification theories propose a quasar nucleus is hidden 
within every radio galaxy.
In the rest-frame ultraviolet, the nucleus is heavily obscured 
along the line-of-sight by 
a dusty torus, but as one observes at longer (infrared) or much shorter (X-ray) 
wavelengths, the optical depth toward the nuclear source decreases significantly.
Therefore, our rest-frame optical NICMOS images may be able to detect nuclear 
point sources where shorter wavelength observations could not.  
To search for possible nuclear point sources we added a delta-function at 
the center of the model galaxies which, upon convolution, becomes the PSF.  
This method is complicated by a degeneracy between the 
central surface-brightness produced by the de Vaucouleurs model and the point-source.  
Also, the galaxy may not follow the \rq-law in the center, and 
instead have a more steeply rising, ``cuspier'' profile which 
would also mimic a central point-source.
A possible resolution to this conflict is to deconvolve the data, 
an effort we have not undertaken at this time.  
The delta-function approach does allow us to separate 
the galaxies into those with and without point sources 
with fairly good accuracy as shown by the simulations (\S\ref{simulations}).  

As a corollary to this method, we can use results from local, 
well-resolved galaxies to potentially further constrain the 
nuclear flux contribution.  
Previous work on low-redshift AGN and ``normal'' ellipticals has 
broadly categorized surface-brightness profiles at small radii \citep*{LAB+95}. 
The profiles are well-described by a ``broken'' power-law.
In general the slope becomes shallower within a ``break'' radius of 
100 -- 300 pc and does not switch sign.  
If the same rule-of-thumb holds true for distant galaxies, then it is 
relatively easy to derive both an upper and lower limit to the contribution 
of a nuclear point source to the galaxy surface-brightness.
At the redshifts of our sources, the PSF spans roughly 600 pc FWHM, so 
we are not able to detect the power-law break directly.  
However, a flux estimate for a central point source component is 
greatly affected by the model behavior in the central portion of the PSF.
If we fit a single power-law and see a central excess in the 
residual image, we have a robust detection of a 
cuspy profile which implies the presence of a nuclear source.  
While this is a sound theoretical argument, in practice 
none of our sources exceed this extrapolated power-law.  This does not 
rule out the presence of point sources, but suggests that 
their luminosities must be relatively low.  
Of the nine galaxies for which we fit full 2D surface-brightness models, 
three of them show detections of nuclear point sources using 
the delta-function approach.  None 
of the three point sources, in 3CR 65, 265, and 280, are brighter than $\approx 5\%$ 
of the total galaxy light. This is consistent with the null result of the power-law method.  
We have calculated robust 1$\sigma$ upper limits to the point-source 
contributions in the other six galaxies by gradually increasing the 
nuclear luminosity while allowing galaxy centroid, $R_e$, and $I_e$ to 
vary, until $\chi^2$ increases to a certain value.  These upper limits are 
included in column 6 of Table~\ref{magtab}.

\subsection{Galaxy Luminosities and Sizes\label{LumMass}}

\begin{figure*}[t]
\plotone{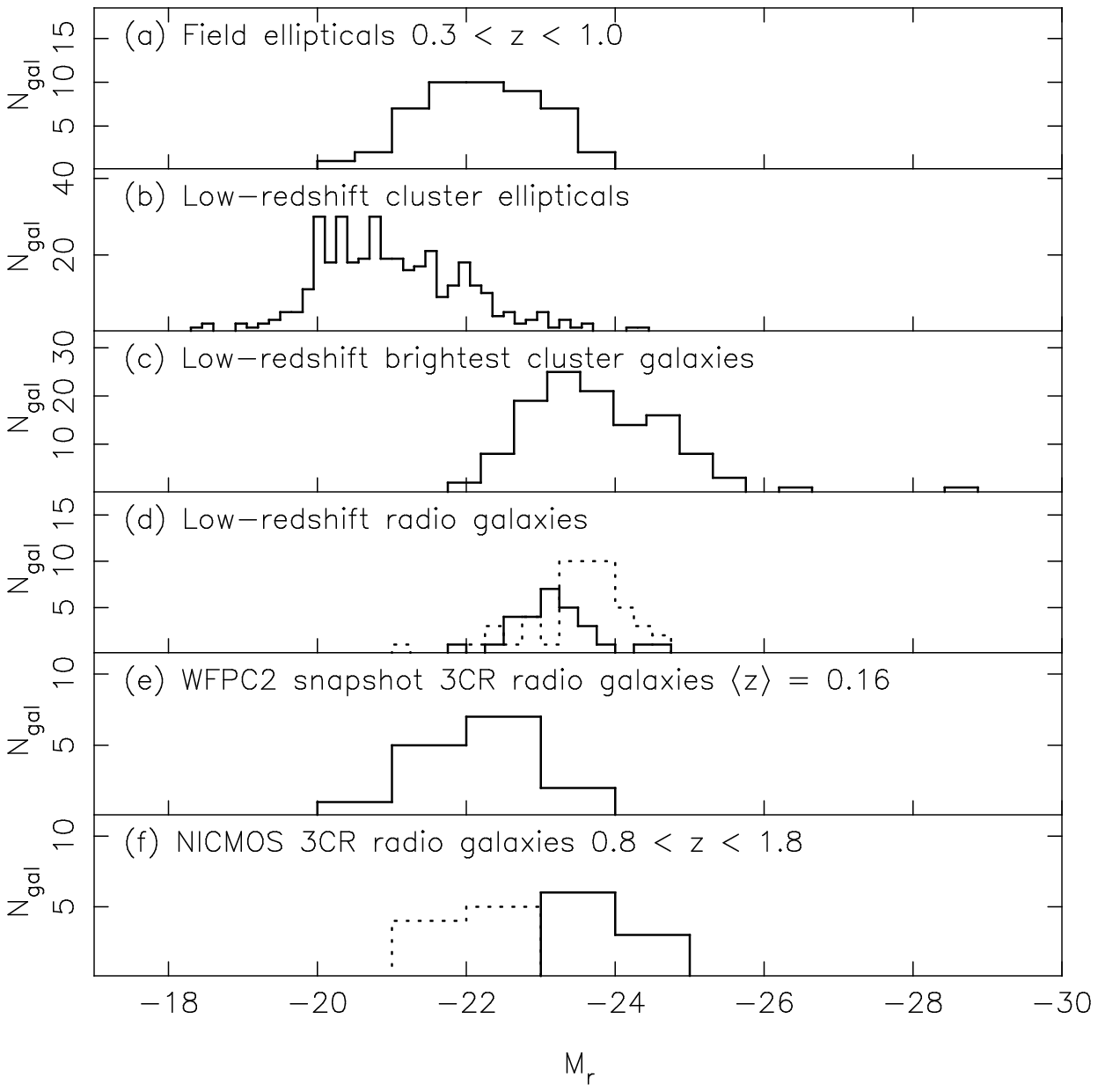}
\caption{Galaxy luminosity distributions for the same samples as Figure~\ref{sizehist}.  
In panel (f) we have included both the observed luminosities of our sample galaxies 
(solid histogram) and those luminosities passively evolved to zero redshift (dotted 
histogram).  The evolved magnitudes overlap the low and intermediate redshift 
radio galaxy distributions with relatively high statistical significance.  
\label{lumhist}}
\end{figure*}

	Using the results of the 2D fits, the galaxy redshifts, and 
an assumed cosmology 
(\S\ref{Intro}), we derive absolute magnitudes in several bands and physical 
sizes from the model fits.  We then compare these 
quantities to those measured in other galaxy samples.

	Table~\ref{magtab} presents the apparent magnitudes of the host galaxy
calculated three ways: 
from the light within a 63.9 kpc metric aperture for the data (Col.~3),
from the light within a 63.9 kpc metric aperture for the model fit (Col.~4), 
and from the total light of the model fit (Col.~5). We also list the magnitudes
of the nuclear point source component (Col.~6).  
These host magnitudes are largely free of contamination from 
AGN-related components. Indeed, we have fit and subtracted the nuclear point sources 
and the aligned light is faint.  
The degree to which these two components do affect the raw data 
(as measured by the aperture magnitudes) can 
be estimated in a mostly model-independent way.  
The point-source component is centered on the peak of the profile and the
absolute upper limit to its contribution is equal to twice the flux 
contained within a diameter equal to the PSF FWHM (because for a gaussian PSF
half of the total flux is contained within the FWHM).
Similarly, most of the aligned light can be measured in the residual image 
where both the host galaxy and point-source have been subtracted away.  
The magnitude of the residuals is usually less than 4\% of the 
total aperture galaxy magnitude.  The one outlier is 3CR~368 which has 
$\approx 14\%$ of the total galactic flux in several bright aligned clumps.  
For all the galaxies the upper limit to the central flux 
is less than 10\% of the aperture flux.  
The total magnitudes quoted in column 5 of Table~\ref{magtab} 
are based on the model fits alone and are used throughout the following analysis.  

	The NICMOS filters, F165M and F160W, cover 
the rest-frame wavelengths near 6500~\AA\ in our sources.  It is 
therefore straightforward to correct our $H_{160}$ measurements 
to the rest-frame Gunn-$r$ band using a model spectral energy 
distribution (SED) as described below.  
We have chosen rest-frame Gunn-$r$ because of the wealth of 
comparison data in this band.  
The choice of SED makes little difference for this small correction, 
but has a significant impact when we attempt to calculate 
magnitudes which are evolved to the present-day.  
Absolute un-evolved $r$ magnitudes are presented in Tables~\ref{partab} 
and \ref{snaptab} for the $z \approx 1$  (NICMOS) and $z \approx 0.16$ 
(WFPC2 snapshot) 3CR radio galaxies respectively.   

	To calculate both rest-frame and passively evolved magnitudes 
we have adopted a model SED from the Bruzual-Charlot (BC) code 
\citep*[a 1999 version of][]{BC93,BRU00}.  
The model SEDs in the BC system are $f_{\lambda}$ spectra, 
normalized to one solar luminosity.  
We have drawn our SEDs from the set of empirical single stellar population models 
with a Salpeter IMF and solar metallicity.  
To calculate the stellar age, we assume that our galaxies 
had their last major epoch of star-formation 
in an instantaneous burst at $z_{f} = 2.5$ and have evolved passively 
to their observed redshift.  This choice of SED and stellar 
age implicitly determines galaxy colors,
which we use to translate our NICMOS observations into other bandpasses,
and mass-to-light ratios, which we use
to calculate rough stellar masses for our sources.

	By integrating the model $f_{\nu}$ SED 
through the appropriate NICMOS filter we normalized the 
spectrum to our observations.  
This normalized SED was then used to calculate 
the color correction to any other desired band.  
To correct to the Gunn-$r$ band above we took the normalized 
SED, integrated it through the rest-frame $r$ band filter 
and applied the AB to Gunn zeropoint offset.  Then to calculate 
the rest-frame Gunn-$r$ from the observed $H_{160}$ we used:
\begin{equation}
r = H_{160} + 2.5 \times \textrm{log}_{10}(\langle f_{160} \rangle/\langle f_{r} \rangle) + 0.23
\end{equation}
where $\langle f_{160} \rangle$ and $\langle f_{r} \rangle$ are the average $f_{\nu}$ 
integrated over each filter and the 0.23 is the zeropoint offset 
going from AB to Gunn-$r$ magnitudes \citep*{FSI95}.  
Applying the same normalization to an evolved SED and calculating the 
corresponding color correction, we derive ``evolved'' magnitudes 
which we use to compare the $z \approx 1$ galaxies with local samples.  

	We have also calculated rest-frame $K$ magnitudes,
included in column~10 of Table~\ref{partab}.  
The more significant color correction from our observed wavelength 
to the rest-frame $K$-band (also included in Table~\ref{partab}) 
depends weakly on the assumed burst redshift 
via the amount of color evolution of the stellar population.  
If we change the redshift of the last major epoch of star-formation 
from 2.5 to 10, the mean absolute $K$ magnitude 
brightens by 0.1 mag.

	Using the normalization and the gas mass fraction of the SED 
we also derive stellar masses.  
As shown in column 11 of Table~\ref{partab} these fall in the range 
of $10^{11} \Msun$.  These derived stellar masses are only 
weakly dependent on the assumed epoch of the last major episode of 
star-formation.  
We feel the choice of passive evolution is justified by both the 
\Kz\ diagram and our later analysis of the size--surface-brightness 
relation (\S\ref{KR}).  

	The galaxy half-light radii are presented in Table~\ref{partab} 
in both arcseconds (Col.~2) and kiloparsecs (Col.~3).  
For consistency with previously published studies, we have only used 
sizes derived using the \rq-law profile rather than those from the Sersic fits.
The galaxies are not perfectly round; the ``radius'' value we have 
quoted is the geometric mean of the semi-major 
and semi-minor axes.  Radii for the $z \approx 1$ 3CR galaxies span the 
range from 3 to 17 kpc.  We find the smallest galaxies also 
happen to be those with the largest 
point source contributions.  We suspect that this is at least 
partly a coincidence, since our simulations show that the radius 
determination is accurate when fitting for the point source. 
	
\subsection{Comparison With Other Samples\label{Comparison}}

\begin{figure*}[t]
\plotone{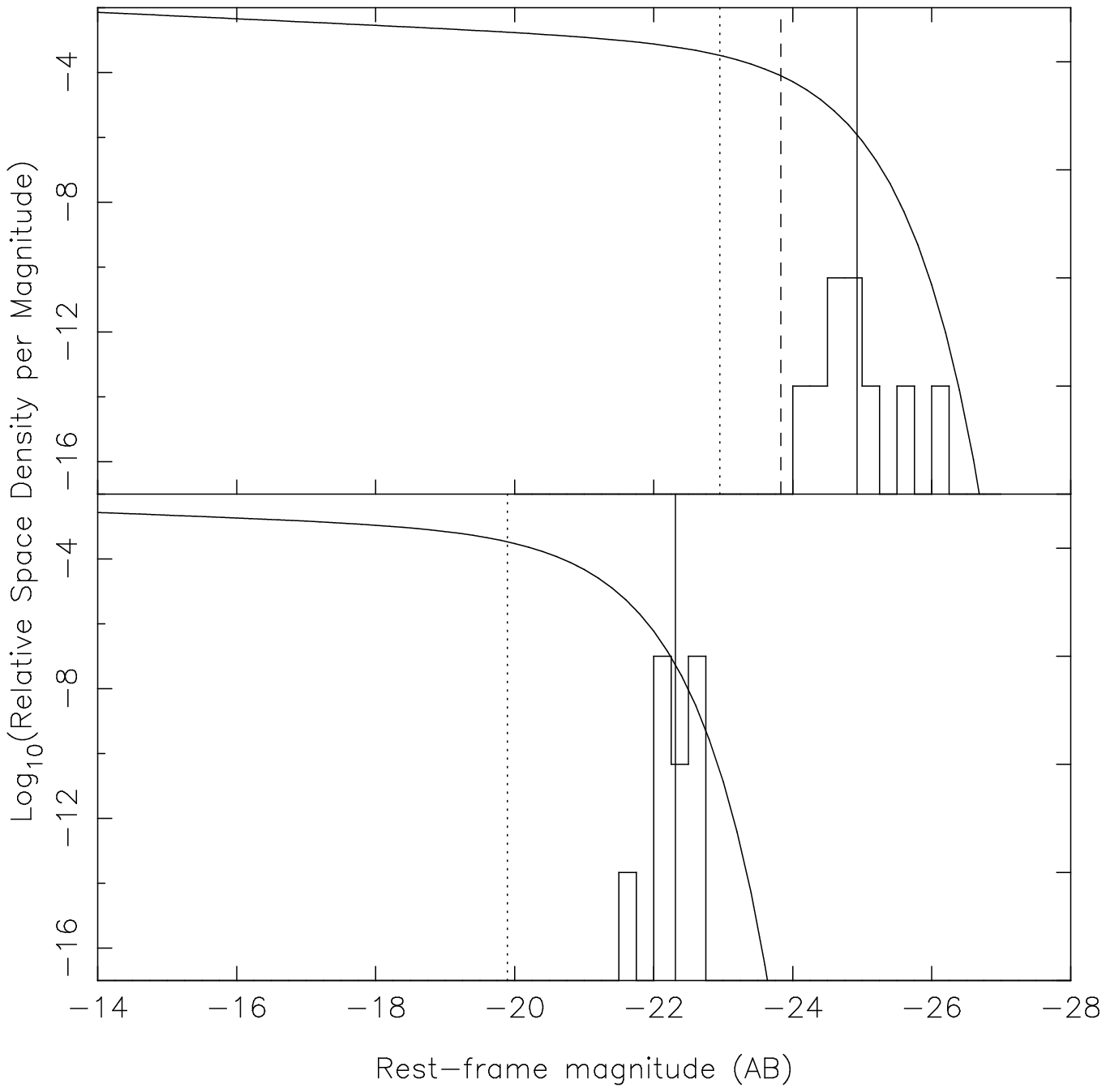}
\caption{Top panel: $K$-band luminosity function from \citet{CSH+96} with arbitrary 
normalization.  The histogram is the observed luminosity distribution 
of the $z \approx 1$ radio galaxies.  The mean RG luminosity (solid line) 
falls at the very luminous end of the ``normal'' galaxy distribution 
($M_{\star}$ for the \citet*{CSH+96} and \citet{dPSE+99} samples are 
shown as the dotted and dashed lines respectively).  
Bottom panel: $r^{\prime}$-band luminosity function from the 
Sloan Digital Sky Survey \citep*{BDE+01} with arbitrary normalization.  
The histogram is now the passively evolved (to zero redshift) 
luminosity distribution of the $z \approx 1$ radio galaxies.  
Again, the evolved sources are brighter than $M_{\star}$ (dotted line) 
suggesting that they are rare, massive sources.
\label{LFfig}}
\end{figure*}

	To put our sample into the context of early-type galaxies in general, we 
compare the $z \approx 1$ 3CR galaxy sizes and luminosities to those 
drawn from previous studies.  The effective radii and host galaxy 
luminosities derived above are directly comparable to previous 
studies of \rq galaxies.  
For our comparison datasets, we have chosen the 
samples of \citet*{SLC+99} (field early-type galaxies 
to $z=1$), \citet*{JFK95} (local cluster ellipticals), \citet*{GLC+96} 
(local brightest cluster galaxies), and \citet*{GFF+00} (local radio galaxies).
We have also included the WFPC2 3CR snapshot 
survey data described above (\S~\ref{Alignment})

The \citet{SLC+99} sample was selected from the CFRS \citep*{CFRS} and LDSS 
\citep*{LDSS} spectroscopic surveys based on 
two-dimensional surface-brightness fits to follow-up 
{\it HST}/WFPC2 F814W imaging.  
Each galaxy included in their analysis was better 
fit by the \rq law than either the 
exponential disk or a combination (disk+bulge)
of the two models.  The resulting catalog contains 48 
elliptical galaxies and ranges in redshift from 0.281 to 0.992.  
We cannot strictly compare the luminosities of this sample 
to our own because it contains galaxies over such a wide range of redshift.  
However, we feel that the more general comparison to such 
a sample has merit since the effective radii are 
still directly comparable to ours and the luminosities, while not 
strictly appropriate, should span the parameter space of all galaxies.  
The \citet{JFK95} study is a collection of data for 211 early-type 
galaxies from 10 different nearby clusters.  Photometry was 
performed on images obtained with the Danish 1.5m telescope 
at ESO.  The targets were selected based on 
previous identification as cluster members.  
Magnitudes and sizes were derived using a one-dimensional 
\rq fit to the photometric curve of growth.  
The \citet{GLC+96} sample utilizes the photometry done in support of 
the  \citet{LP94} study of peculiar velocities 
of Abell clusters.  The \rq 
fits to the BCGs 
were performed on isophotal profiles extracted 
from the ground-based images.  The \citet{GFF+00} sample is 
selected from two radio flux-limited samples \citep*{WP85,EWS+89}.  
Galaxy sizes and surface-brightnesses were derived via 
2D image modeling of ground-based $R$-band images of 
79 low-redshift ($z<0.12$) FRI and FRII radio 
galaxies.  
The authors include three components in their model fits, an \rq profile, 
a nuclear point source (represented by the PSF) and an 
exponential disk.  The three components were 
combined to obtain the optimal fit.  For the 
size comparisons presented here we use the \citet{GFF+00}
empirically determined half-light radius. For the luminosity comparison 
we have adopted their quoted total magnitudes, which are 
based on an extrapolation of the \emph{observed} surface-brightness 
profiles rather than the model fits. 
However, for the purpose of constructing the Kormendy relation we 
have used solely the parameters from the de Vaucouleurs fit.  
It should be noted that the Govoni et al. RGs (particularly the 
FRIs) have total magnitudes (which include point-source and 
disk components) and sizes comparable to those of the BCGs.
The galaxy luminosities and sizes of these samples have all been 
converted to our chosen cosmology.

\subsubsection{Sizes}

\begin{figure*}[t]
\epsscale{1.5}
\plotone{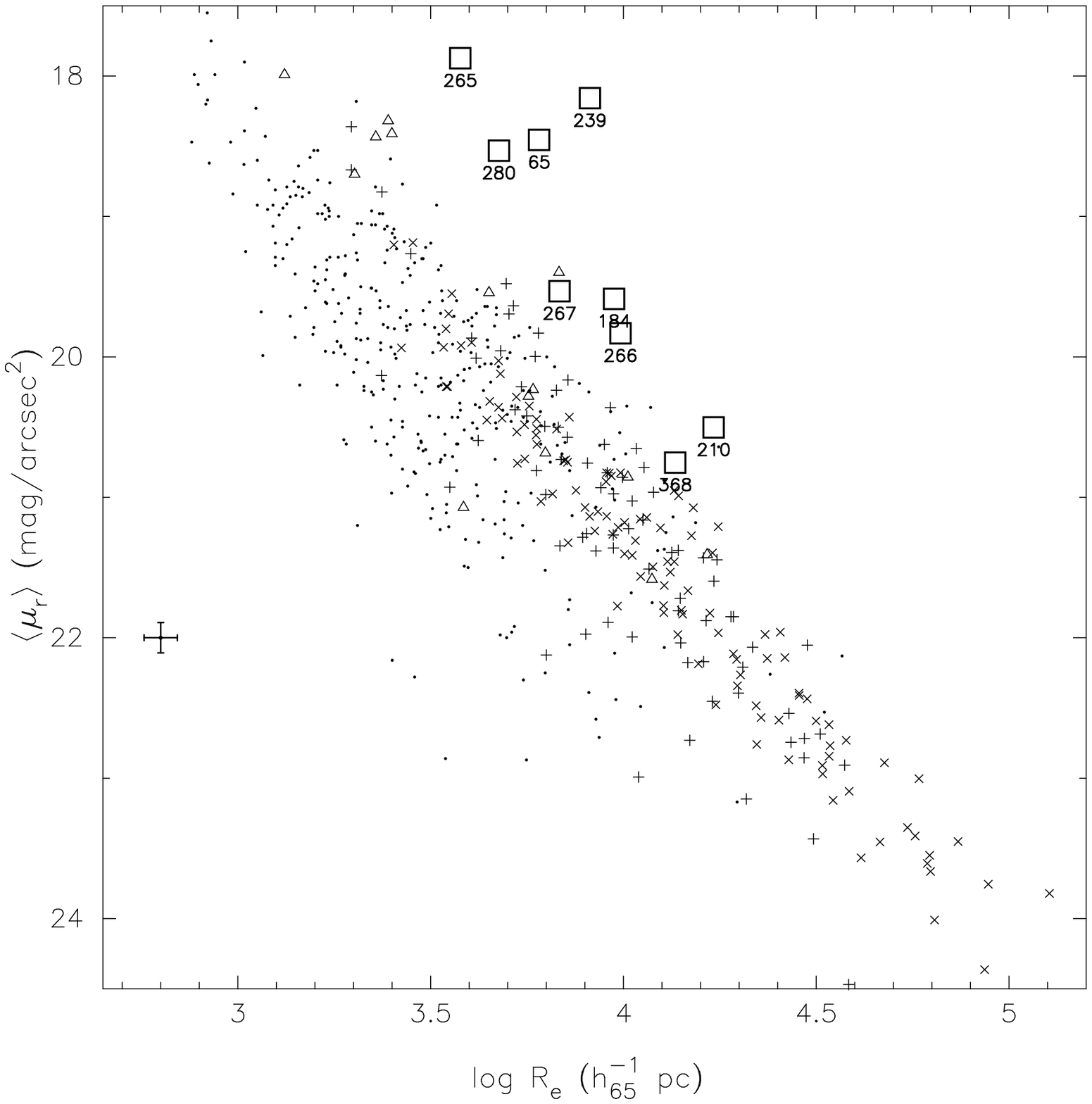}
\caption{Kormendy relation in rest-frame Gunn-$r$.  The dots are cluster ellipticals 
from \citet*{JFK95}.  The pluses are the low-redshift galaxy 
sample of \citet*{GFF+00}.  The triangles are intermediate redshift ($z_{{\rm median}} = 
0.16$) 3CR radio galaxies from the WFPC2 snapshot survey \citep{MBS+99}, fit with our code.  
The crosses are the brightest cluster galaxies from \citet{GLC+96}.  
The large squares are the NICMOS 3CR data and are labeled by name.
A representative average error-bar for the NICMOS data 
is shown in the lower left panel.\label{krfig}}
\end{figure*}

In Figure~\ref{sizehist} we present the size distributions 
from our own sample along with the data from the literature 
described above.  The NICMOS galaxies have a wide 
size distribution spanning $3.7$ to $17.1$ kpc.
To the eye, their sizes seem most similar to the FRII galaxies 
of the \citet{GFF+00} sample (solid line in Fig.~\ref{sizehist}).  
We employ the KS test to statistically compare the $z \approx 1$ 3CR galaxies
to all the other samples.
The local cluster, brightest cluster, and intermediate-redshift 
field galaxies are wholly inconsistent 
with the NICMOS 3CR size distribution at greater than 99.5\%
confidence.  The most consistent score, at a statistically 
inconclusive 65\%, 
is found for the local FRII galaxies as suggested by inspection.  

There is no exact 
counterpart population at low redshift.  
Lacking such a direct correspondence, we assert 
that it is unlikely that galaxy size decreases 
with time.  Mergers will either leave the 
size unchanged, as in the case of accretion of a low-mass galaxy, 
or perhaps increase the size, in the case of equal mass mergers (e.g., 
\citet*{CdCC95}). 
If the merger funnels gas to the galaxy 
center, where it then forms stars, the merger may temporarily 
shrink the half-light radius.  However, this effect will 
not dominate the observed light for long and is therefore 
not likely to be a major contributor to observed galaxy properties.  
Under the scenario where their sizes either remain the 
same or increase slightly,
the $z \approx 1$ 3CR galaxies match up 
with the local RGs or, if they grow, 
overlap the BCGs.  

The one outlier at small size (3CR~332 at $\approx 0.3$ kpc) 
in the WFPC2 snapshot sample is contaminated 
by a strong point source.  This source has 
a doubly-peaked $H\alpha$ line as shown by \citet*{HALP90}, 
suggesting that we may be observing the nucleus/accretion-disk more directly 
than in most other radio galaxies.  
The derived nuclear luminosity in this case is much 
higher relative to the host galaxy than for the NICMOS galaxies.  
If this galaxy is removed, the intermediate-redshift 
radio galaxy size distribution becomes more consistent with the 
high-redshift one.

\subsubsection{Luminosities\label{Lum}}

In Figure~\ref{lumhist} we present the 
luminosity distributions for the NICMOS sample 
along with the previously published samples 
described above.  
Like the size distributions, 
the $z \approx 1$ radio galaxy luminosities are roughly bracketed 
by the luminosities of the other samples.  
Again, we use the KS test to compare the distribution of 
our sample galaxies to those from the literature.  
In contrast to the sizes, however, we perform this comparison 
for the observed values directly, but also for the observed values
modified using an assumed evolutionary scenario.

As determined by the KS test,
the high-redshift radio galaxies are not drawn from 
the low-redshift cluster elliptical or intermediate-redshift 
radio galaxy distributions at greater than 99.9\% confidence. 
The exceptions are the BCG sample and 
the local FRI sample.
The value for the BCG comparison is a statistically 
indeterminate 60\%, meaning that the $z \approx 1$ 3CR 
sample is inconsistent with being drawn from the 
BCG sample at this significance.
The value for the low-redshift FRI sample is an also 
inconclusive, but slightly more consistent, 20\%.
It is interesting to note that while the sizes seem 
most consistent with the low-redshift FRIIs, the 
luminosities coincide with the local FRIs.  
In order to make a more direct comparison with local samples 
we have evolved the $z \approx 1$ 3CR galaxies to 
$z = 0$ using the method outlined above (\S~\ref{LumMass}).

When we do this, the
luminosity distribution for the current sample shifts fainter 
by 1.5 - 2.0 magnitudes via passive evolution 
to zero redshift
(dotted line in panel f of Figure~\ref{lumhist}).  
The amount of evolution was calculated using 
the normalized model SED described in \S~\ref{LumMass}.  
We applied the same normalization to the model 
spectrum aged to zero redshift, and integrated 
it through the rest-frame Gunn-$r$ filter.  
With this sizable correction the magnitudes are 
again bracketed by the low-redshift samples.  
Now the BCGs, cluster galaxies, and local RGs 
are rejected at $> 99\%$ significance.  
While no sample is exactly comparable, the 
intermediate-$z$ RGs 
are only rejected at the 34\% significance level.
If 3CR~332 is again thrown out on the basis of its 
discrepant size, the luminosity distribution of the 
intermediate-redshift RGs is rejected at only the 
25\% level, making it the most consistent of the 
comparison samples.

Another perspective can be taken by comparing 
the luminosities with 
much broader, more statistically robust samples.  We can 
judge the relative rarity of the radio galaxies 
when compared to the more general luminosity functions (LFs) 
derived from large galaxy samples.  
At low redshift, the Sloan 
Digital Sky Survey provides a large number of galaxies 
for which the luminosity functions have been computed \citep*{BDE+01}.  
At higher redshift ($0.6 < z < 1.0$), 
we compared our 3CR galaxies to the field galaxy LF from \citet*{CSH+96} 
and the cluster galaxy LF from \citet{dPSE+99}.
The \citet*{CSH+96} spectroscopic targets were chosen from 
an infrared imaging survey 
covering $26.2$ $\square^{\prime}$ of the Hawaii 
Deep Field (areas SSA 13 and 22).  
We use their measurement of $M_{K}^{\star}$ and $\alpha$ for the $0.6 < z < 1.0$ 
bin as a comparison to our sample.  
The more recent study done by \citet*{dPSE+99} investigates the 
$K$-band LF in galaxy clusters to $z \approx 1$.  
In Figure~\ref{LFfig} we show a histogram of 
our galaxies along with the LFs from the 
\citet*{CSH+96}, \citet{dPSE+99}, and \citep*{BDE+01} samples.  
Using the assumed passive evolution model as 
above, we have calculated both the rest-frame 
$K$-band (with no evolution) and $r$-band 
magnitudes (evolved to zero redshift) to compare with these two LFs.  
In both cases, the HzRGs are at the very luminous end 
of the LF, underscoring their relative rarity.  

\subsubsection{Kormendy Relation\label{KR}}

Local spheroids (early-type galaxies and spiral bulges) 
define a continuous and low-scatter 
sequence in the size--surface-brightness plane.  
This two-dimensional projection of the three-dimensional 
Fundamental Plane (FP) provides a useful 
diagnostic of galaxy populations 
without velocity dispersion measurements.    
While the physical processes responsible for the correlation
are not entirely understood, a scaling relation 
of this sort is predicted qualitatively by the 
application of the virial theorem to these stellar 
systems (assuming a single 
mass-to-light ratio).  
Although this may only be illustrative, a galaxy which lies on the Kormendy 
relation is likely to be in dynamical equilibrium 
and may share a similar formation history with the other 
galaxies on the size--surface-brightness locus.  

Figure~\ref{krfig} shows our sample 
alongside the same low-redshift comparison galaxies, 
excluding the intermediate redshift field ellipticals 
from \citet{SLC+99}.  Of these samples, the local cluster, 
brightest cluster, and radio galaxies are known 
to lie on the FP \citep*{JFK95,OH91,BFF+01}.  
The 3CR snapshot galaxies form a tight sequence
in the size--surface-brightness plane, 
but kinematic measurements are not yet available 
for their placement on the FP.  
A representative $1\sigma$ error bar for the $z \approx 1$ 3CR galaxies, 
based on our error analysis in \S~\ref{simulations}, 
is shown in the lower left corner of the figure.  

The $z \approx 1$ 3CR galaxies form a narrow sequence in 
the size--surface-brightness plot.  This is not expected a priori for a 
set of sources selected via 
AGN rather than stellar properties (i.e., radio flux).  
To test the significance of this correlation defined 
by the high-redshift 3CR 
galaxies, we have employed a simple rank correlation 
analysis.  The small value of the two-sided significance 
indicates a strong correlation.  
The slope of the Kormendy relation is well-defined for low-redshift 
galaxies where evolutionary effects within the galaxy samples 
are small.  For the NICMOS sources presented here 
this is not the case; several gigayears ($=3.3$ for 
our cosmology) have passed 
between the observed epochs at $z=1.8$ and $z=0.8$.  
To alleviate this systematic bias we have evolved 
our sources to a common redshift, $z=0$, 
following the procedure outlined above 
and assuming no evolution of galaxy sizes.

\begin{figure*}[t]
\plotone{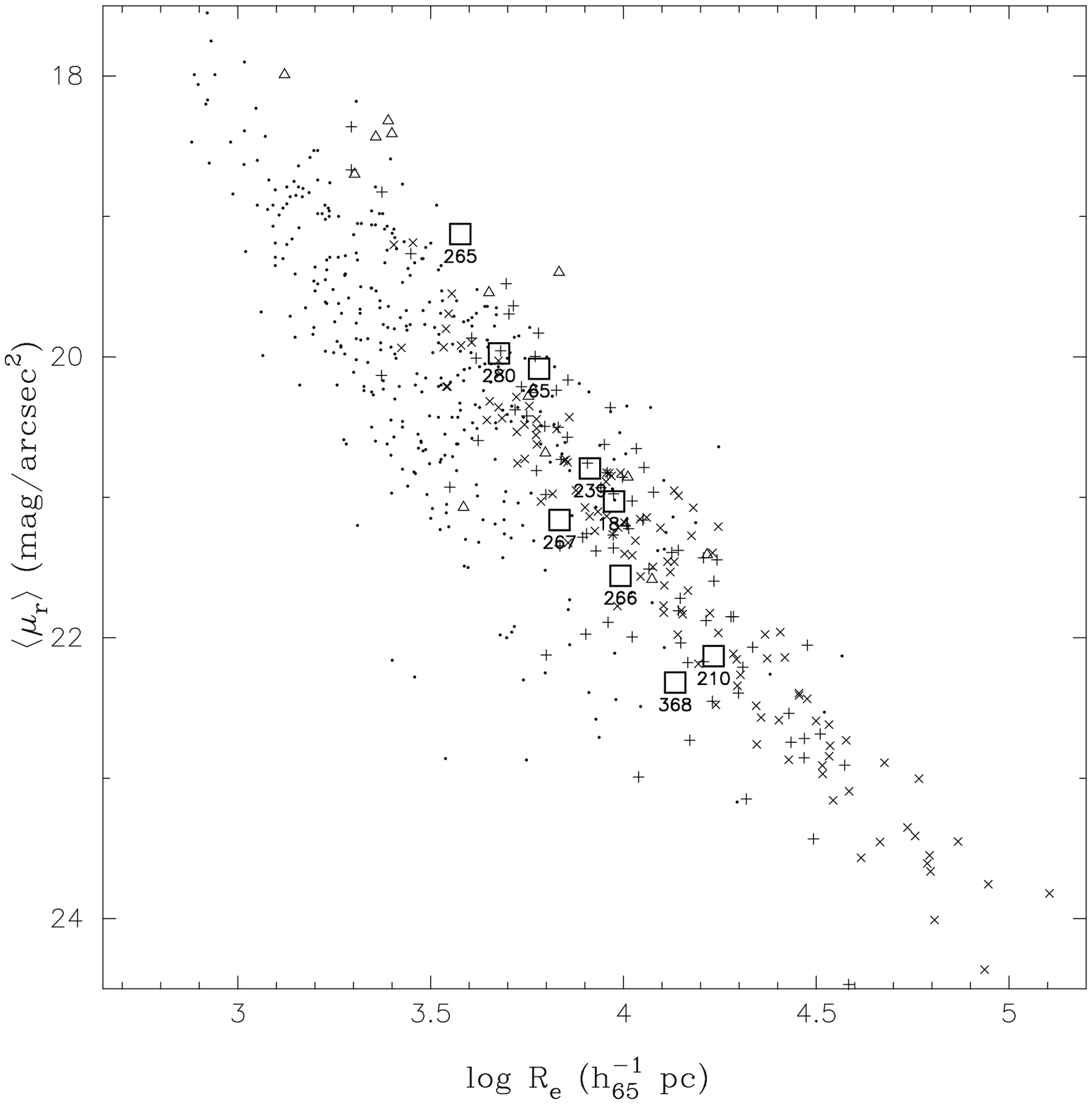}
\caption{Same as Figure~\ref{krfig} except the NICMOS galaxies have been passively evolved
to zero redshift using the stellar population model ($z_{f} = 2.5$) 
described in the text.  The $z \approx 1$ 3CR galaxies evolve 
roughly onto the local relation, but are not statistically consistent with the 
distributions of the comparison samples.
\label{krevolfig}}
\end{figure*}

Figure~\ref{krevolfig} again shows the same comparison galaxies 
except with the NICMOS sources now 
passively evolved to zero redshift.  The differential evolution between 
the samples allows the $z \approx 1$ galaxies 
to fall roughly on the local relation.  
The slope of the HzRG relation is steeper ($\approx 4.7$) than that 
of the local galaxies ($\approx 3.5$).  If we throw out the three 
point source galaxies (3CRs 65, 265, and 280) the slope is much shallower 
($\approx 3.6$) and is more in line with the local value.  
\citet{BLR98} and \citet{McD00} correspondingly derive slopes 
of $5$ and $3.5$ for their similar sample of $z \approx 0.8$ RGs.  
It is interesting to note that our values range between these 
two slopes dependent upon the inclusion of the galaxies with 
a nuclear point-source.  
The slope is known to be very sensitive, particularly in 
these small samples, to observational selection effects.  
To test more robustly whether the evolved sources do, in fact, align 
with the low-redshift relation, we have utilized the 
two-dimensional KS test \citep*{P83,FF87}.  This 
test rejects the hypothesis that the $z \approx 1$ 3CRs 
(evolved to zero redshift) are  
drawn from the same distribution as 
any of the local samples, at relatively high 
significance ($81\%$ for BCGs to $99.9\%$ for cluster members).  This 
does not contradict the observed overlap between the high- and low-redshift 
samples, but rather shows that they are not statistically identical 
(e.g., as we have already shown, the size and luminosity distributions of the 
NICMOS 3CR galaxies differ from that of any particular comparison sample).   
A more lenient test is to look at the surface-brightness distributions 
in the size range of the HzRGs.  
Using the one-dimensional KS test for these distributions, we find
that the evolved 3CR galaxies are most similar to the low-z FRIs, 
followed by (in order of increasing rejection)
the intermediate-redshift RGs (from the WFPC2 fitting), 
the low-z FRIIs, and the low-z cluster galaxies.
In any case, the HzRG surface-brightnesses overlap those 
for galaxies of similar size at low redshift when a simple 
model of passive luminosity evolution is applied.  

\subsection{Radio Galaxy Evolution}

	The NICMOS data presented here are unique 
in their combination of high spatial resolution and 
near-infrared wavelength coverage.  These capabilities 
have allowed us to study the properties of $z \approx 1$ 
radio host galaxies in some detail.  
The RG sizes, luminosities, and morphologies all have 
implications for the complete picture of radio, as well as
perhaps ``normal'', galaxy formation and evolution.

	A single, passively evolving population which formed 
at early times ($z \simgt 2.5$) fits the observed 
\Kz\ relation for HzRGs very well \citep*[e.g., ][]{LL84, JRE+01}.  
This result has defined the current radio host 
evolution paradigm: a massive, passively evolving 
stellar host dominates the 
rest-frame optical light and is responsible for the \Kz\ relation, 
and a UV bright, but dynamically insignificant,
component accounts for the alignment effect.  
Our data focus on the putative stellar component.  
Are the rest-frame optical morphologies, sizes, and luminosities
consistent with the passive evolution picture?

The existence of a luminous, apparently stellar,  
continuum component
undetected in the rest-frame UV supports the 
paradigm at its most basic level.  
Furthermore, the galaxy morphologies 
are well-described by the de Vaucouleurs profile,
which at low redshift is associated with old, red elliptical galaxies.  
Ellipticals in clusters and the field 
seem to be passively evolving from early formation times \citep*[e.g., ][]{SED98,SLC+99}.  
In this context, the fact that the HzRGs are well-fit 
by the de Vaucouleurs profile suggests that they 
belong to a similar, if not the same, class of passively evolving sources.  

The passive scenario also dictates that 
the radio hosts will undergo little or no size evolution.
The HzRGs have similar effective radii 
to the low-redshift samples, again supporting the paradigm: 
the sizes of the $z \approx 1$ RGs are most consistent with the 
$z \approx 0$ FRII radio hosts.  
We may, therefore, identify the HzRGs as the progenitors of 
local radio galaxies.  Radio galaxies would then 
be drawn from a single, passively evolving population 
of sources formed at a common high redshift.  
The relative rarity of RGs at any given epoch is 
explained by noting that 
the radio emission in FRII sources is likely to be 
a transient phenomenon occurring on timescales 
of roughly $10^7$ years \citep*[e.g.,][]{AL87,MKO+98,KA99}.  
To follow this argument further, we now discuss the 
luminosities of the HzRGs.  

With no evolution, the $z \approx 1$ radio galaxies
are much more luminous than all but the BCGs and 
BCG-like low-redshift FRI RGs.  
This is not a completely fair comparison since BCGs are perhaps
the product of a ``special'' formation history \citep*[e.g.,][]{WEST94,ABK98,BCM00} 
and are known to have distinct size and luminosity distributions.
If instead we restrict the 
size range of the comparison samples to the same set of 
values spanned by the high-redshift 3CRs, then the 
$z \approx 1$ sample outshines all other galaxies including the BCGs.  
This discrepancy could be due to inherently more luminous galaxies
or to the AGN-related light boosting the derived values.  
Based on the fits and residual images, we believe that the 
AGN contribution cannot account for a majority of the luminosity difference.  
We therefore conclude that the $z \approx 1$ RGs are inherently more 
luminous.  
If we assume they dim passively, then they fall in the range 
of local galaxy luminosities.  
Imposing the galaxy radius restriction 
as above, the high-$z$ 3CR galaxies are fainter than the 
BCGs and low-redshift RGs, and are brighter than the local cluster galaxies.  
In partial support of the paradigm, they line up most 
closely with the intermediate-redshift 
radio hosts.

In the size--surface-brightness plane the HzRGs evolve onto 
the local relation, although distinctly 
distributed from any of the comparison samples.  
Again the HzRG distribution is most consistent with 
the local RGs.  This correspondence, like the evolved luminosities above, 
depends on an assumed star-formation history.  
In order to fade these sources onto 
the local Kormendy relation and into the luminosity 
range of local galaxies using 
the single-burst model, the most recent star-formation
cannot occur too early, roughly $z_{f} \approx 2-2.5$. 
However, most of our galaxies would evolve 
sufficiently with $z_{f} \simlt 10$.  
It is the highest redshift source in our fit 
sample, 3CR~239 at $z=1.78$, that drives to lower redshift the choice 
for the last major star-formation (which 
is not necessarily the same as the last major merging epoch).  
For a galaxy at $z = 1.8$, changing the ``formation'' redshift from 2.5 
to 5 decreases the amount of fading to $z = 0$ by about 0.8 
magnitudes. For the other, lower redshift,
galaxies in our sample the 
difference is smaller, about 0.3 magnitudes.  
Such a low ``formation'' redshift to accommodate 3CR~239 
is problematic because the RG \Kz\ relation 
is known to extend to $z \simgt 5$ with low scatter.
In addition, 3CR~239 exhibits several discrete clumps 
of emission surrounding the central concentration, 
suggesting possible residue of formation or 
in-falling satellites.  
It is also interesting to note that the 
highest redshift object in the entire sample, 
3CR~256 at $z=1.82$, shows no centrally 
peaked emission and is underluminous at $K$ 
compared to the \Kz\ relation  
at that redshift \citep{CE91,SEA+99}.  
While two data points do not constitute a 
robust sample, these galaxies suggest that 
the epoch for radio galaxy formation may 
extend in some cases to $z \simlt 2$.  
At $z \simgt 2$ it has been shown that 
a majority of RGs, while falling on the $K$-band Hubble 
diagram, show few signs of smooth, round 
host galaxies \citep*{vBSS+98,PMcR+01}.  
So, either we maintain that passive evolution 
is consistent with our data and 3CR~239 is 
an outlier (although, unlike 3CR~256, it does not deviate significantly 
from the mean \Kz\ relation),
or we must imagine an alternate explanation for 
the observed Hubble diagram.  

If the distant radio hosts do not evolve into 
the local radio-loud galaxies, how is the 
\Kz\ relation explained?
An alternative posited by \citet{BLR98} is that low- and 
high-redshift RGs are not linked by evolution, but rather 
that RGs are only created in host galaxies above a given 
large galaxy mass.  
Their main supporting evidence for this view was that 
the cluster environments of RGs seemed to become less 
rich at lower redshift, something that cannot happen.  
The inclusion of a galaxy mass cutoff makes sense in terms
of the relationship which has been discovered 
at low-redshift between black-hole mass 
and bulge velocity dispersion 
or luminosity \citep{MTR+98,GBB+00,FM00}.  
In this scenario, the natural relation between 
radio luminosity and redshift in a flux-limited sample 
populates the Hubble diagram in a smooth, continuous way 
via the correlation of radio power with black hole and 
bulge mass.  
While our data cannot address the environment issue, 
we do not see any comparable ``impossible'' transformation 
(e.g., low-$z$ RGs being smaller than HzRGs or fainter than 
passive evolution would predict)
that precludes radio galaxies at high-$z$ from being 
drawn from the same population as low-$z$ RGs.
Given our data, the simplest 
explanation of the \Kz\ diagram, passive evolution, 
remains tenable.

The fact that RGs form a highly-correlated group and follow largely the same 
analytical profiles (with possibly interesting differences) 
suggests that they may form a homogeneous sample.  
Similar behavior among local galaxy samples has been used to argue that 
the constituent galaxies have followed similar evolutionary paths.  
While our sample is too small to make such grandiose statements, the 
size--surface-brightness
data certainly do not preclude the possibility that this radio-selected 
sample constitutes a homogeneous subset of the entire galaxy population.  
This is seemingly not true for the higher redshift objects, 
3CR~239 and 256 in our sample and the $z \simgt 2$ galaxies of 
\citet{vBSS+98} and \citet{PMcR+01}, which begin 
to show interesting diversity in their properties.  

\section{Conclusions and Future\label{Conclusions}}

	Historically, there have been many attempts to 
unify the disparate qualities of high-redshift radio galaxies 
into a full picture of their formation and evolution.  
{\it HST} and NICMOS now play a crucial role in this continuing story.  
	The present observations have targeted 11 $z \approx 1$ 
3CR radio sources for detailed study.  Round, smooth galaxies appear 
in the NIR and outshine the complex morphologies seen in the rest-frame UV.  
The NICMOS data have allowed us to analyze these sources in ways that 
were previously impossible or at least ill-advised.

	The salient results of our investigation are briefly summarized:
The majority of our sample are well-fit by the de Vaucouleurs profile, 
with low-level residuals that qualitatively match the rest-frame UV 
morphologies observed with WFPC2.  
The derived surface-brightnesses and sizes match well with bright local 
ellipticals, but are both fainter and smaller than the brightest galaxies 
found in rich clusters at low-redshift.  
Apart from one fit source (3CR~239), passive evolution from 
an early formation epoch ($z \simgt 5$) is sufficient to 
fade the sources to match the local samples.  
The slope of the Kormendy relation for our sample is 
steeper than that found for local galaxies, 4.7 compared to 
3.5, and may be consistent with a constant luminosity line.  
If a lower ``formation'' redshift is chosen, $z = 2.5$, 3CR~239 can 
simultaneously be made to fade into the range of low-redshift galaxies.  
However, comparison to luminosity functions at both low and high 
redshift illustrates the relative rarity of galaxies 
at these luminosities, and suggests that HzRGs are 
quite massive.  
The $z \approx 1$ 3CR host galaxies may show a tendency to 
align their major axes with the axis defined by the double 
radio lobes, but confirmation of this result awaits detailed 
multi-band analysis.

In short, the $z \approx 1$ radio galaxies seem to be rare, massive objects  
which are dynamically relaxed.
As such they are 
likely the progenitors of local giant elliptical galaxies.  
These sources are ideal for follow-up NIR spectroscopy on 
large telescopes.  They are relatively bright at $H$ and should 
have measurable velocity dispersions at relatively low spectral 
resolution.  In addition, similar studies of higher redshift 
RGs will test whether the evolutionary trends seen here between 
$z=0$ and $z\approx 1$ continue to even earlier epochs.   

\appendix
\section{Notes on Individual Objects}
\label{Notes}
{\bf 3CR~265} ($z = 0.811$):  This galaxy lies in a moderately rich
cluster of galaxies \citep*{DELTORNPHD} and is surrounded by spectacularly
extended UV emission seen in the WFPC2 images.  The extended UV light is
strongly polarized \citep*{JE91} and shows extended, broad
MgII emission \citep*{DS96}, implying that it is dominated
by scattered AGN light.  It is, however, significantly misaligned from the 
radio source axis.   The NICMOS image is wholly dominated by a round,
symmetric giant elliptical host galaxy.  The central surface-brightness 
is remarkably high, and the model subtraction shows extended, 
irregular residuals, harboring an obscured point source.  The point source fit 
finds a central nucleus of $m_{pt,160} = 22.84$ in the F160W filter.

{\bf 3CR~184} ($z = 0.994$):  Another cluster galaxy \citep*{DLeFC+97},
3CR~184 is elongated, bimodal, and aligned in the rest-frame UV, but has 
a dominant gE host galaxy in the NICMOS image.  The isophotal elongation of the NICMOS image 
along the radio source axis is most likely due to the low-level, aligned light seen in the 
residuals.  The WFPC2 image here is from \citet{DLeFC+97}.

{\bf 3CR~280} ($z = 0.996$):  The complex aligned structure visible in 
the WFPC2 images of \citet{BLR97} largely disappears in the NICMOS
data, although the highest surface brightness components remain barely
visible.  The radio galaxy host appears to be a smooth, symmetric
giant elliptical, with a central point source ($m_{pt,160} = 23.03$).

{\bf 3CR~368} ($z = 1.131$): The rest-frame UV light from this object
is extended over $7^{\prime\prime}$ in length along the
radio source axis and, surprisingly, is unpolarized.  There is
a galactic M-star superimposed on the galaxy \citep*{HPLeF91},
with another, fainter star at the north end visible in the NICMOS
image.  Both stars were subtracted before performing the surface-brightness 
fits.  
The host galaxy nucleus seen by NICMOS lies at a local minimum
in the WFPC2 images;  here and in other radio galaxies it is likely
that a combination of dust-obscuration and $k$-correction are
responsible for rendering the stellar nucleus completely invisible at
optical wavelengths.  The near-IR central surface brightness of 3CR~368
is low and its surface brightness profile is unusually extended and
``flat''.  It may be that dust obscuration 
remains significant in this object even at 0.8$\mu$m in the rest frame.  
It is also possible, based on the Sersic law fits, that 
the host galaxy of 3CR~368 has a more BCG type profile ($n>4$).  

{\bf 3CR~267} ($z = 1.140$):  The WFPC2 images of 3CR~267 show diffuse
emission, several bright knots, and a broad ``arc'' all distributed generally
along the radio source axis.  In the NICMOS image, the three brightest 
knots appear to be three distinct galaxies:  the gE radio source host
plus two fainter companions, one evidently a disk galaxy.  3CR~267 is 
also known to inhabit a rich cluster.  The fit residuals are very small 
in this case, without including any point source component.  

{\bf 3CR~210} ($z = 1.169$):  The complex, multicomponent structure seen
in the WFPC2 data simplifies to a dominant, if not completely smooth,
central host galaxy in the NICMOS image, together with another red
companion object.  Careful comparison of the WFPC2 and NICMOS data
strongly suggests the presence of a thick dust lane obscuring the 
central regions in the rest-frame UV.  Extinction may be responsible
for the central distortions still seen at the 0.8$\mu$m rest frame in 
the NICMOS data.  The 3CR~210 host galaxy seems to have two distinct 
peaks, which are less distinct in the NICMOS data than in the rest-frame 
UV.  This source also lies near the center of a moderately rich cluster (J-M. Deltorn, 
private communication).  

{\bf 3CR~65} ($z = 1.176$):  This galaxy was long believed to be a giant
elliptical on the basis of ground-based near-IR images 
\citep*[cf. ][]{RL94} and spectroscopy of the 4000\AA\ region
\citep*{SKR95}, although \citet*{LRE+95} have argued
that it harbors an optically-obscured AGN point source which is detectable 
in the near-IR.  Both points of view may be true:  viewed with NICMOS the 
galaxy appears to be a round gE, but with extremely high central surface 
brightness.  The fits find a central point source with $m_{pt,165} = 22.75$.  

{\bf 3CR~266} ($z = 1.275$):  One of the most elongated of all 3CR galaxies
in the rest-frame UV, the alignment of 3CR~266 persists into the 
near-IR, although the isophotes are generally rounder.  
The near-IR morphology is striking; cleanly separating into a relaxed, 
elliptical host and a thin, aligned component nearly identical to the WFPC2 
image.

{\bf 3CR~470} ($z = 1.653$):  This galaxy is extremely faint at optical
wavelengths and is barely visible in the WFPC2 images.  Indeed it is not
entirely clear which of the faint blobs in the WFPC2 data is the 
actual host:  \citet{BLR97} place the radio nucleus between the
two.   Eisenhardt et al.\ (1989) noted that the near-IR structure of
3CR~470 is apparently oriented perpendicular to the radio source axis.
In the NICMOS images it appears that this is due to the (chance?) 
alignment of three independent objects distributed roughly in a line.

{\bf 3CR~239} ($z = 1.781$):  This galaxy is compact and bright in both
the optical WFPC2 and NIR NICMOS images.  The NICMOS image is 
dominated by a single, round, centrally concentrated galaxy, but its outer 
envelope breaks up into several ``satellite'' lumps.  The image is 
suggestive of a giant but not fully mature elliptical galaxy at 
$z \approx 1.78$, perhaps seen in the process of assembling through
accretion.  While the profile is well fit by an \rq-law, the residuals again 
show that extinction may play a large role at rest wavelengths of $0.6 \mu$m.

{\bf 3CR~256} ($z = 1.819$):  The most distant galaxy in our sample, 
3CR 256 is extremely bright in the rest-frame UV and consists of two 
dominant components, strongly polarized, and closely aligned with the 
radio source \citep*{JES+95,DCvB+96}.   
It is elongated/aligned in the near-infrared, and is very subluminous, 
lying well below the mean $K$-$z$ relation at that redshift.  
On this basis, Eisenhardt \& Dickinson (1992) suggested that 3CR~256 
may be one of the few true examples of a protogalaxy among the 
3CR radio sources (see \citep*{SEA+99} for a detailed analysis 
of the spectral energy distribution).  
In our NICMOS F165M images the galaxy is diffuse, elongated and very faint:  
it is the one complete exception to the trend toward compactness and 
symmetry among the NICMOS 3CR observations presented here.  
One peak of the infrared light lies between the bright UV lobes;  
there is also a faint, compact red object barely visible 
in the optical data.  An F160W NICMOS image (not shown here) which 
includes [OIII] line emission has a completely different appearance, 
dominated by two bright lobes corresponding to those seen in the
optical continuum and Lyman~$\alpha$.

\acknowledgments

We gratefully acknowledge M. Giavalisco for providing
his NICMOS imaging data to aid in our calibrations, 
A. Martel for providing calibrated WFPC2 frames from the 
3CR Snapshot Survey, and to B. Jannuzi for allowing us 
to present his WFPC2 image of 3CR~256.  
We also thank Pat McCarthy, Tod Lauer, Adam Stanford and,
Peter Eisenhardt for helpful discussions, and 
the anonymous referee for many 
useful comments that improved both the content and 
presentation of this research.  
This work was supported by HST grants GO-07454.01-A and 
GO-06583.03-A.  Support for proposal \#7454 was provided by 
NASA through a grant from the Space Telescope Science Institute, 
which is operated by the Association of Universities for 
Research in Astronomy, Inc., under NASA contract NAS 5-26555.


\clearpage

\begin{deluxetable}{lllllllllll}
\tabletypesize{\tiny}
\tablewidth{0pt}
\tablecaption{Observation Log}\label{obstab}
\tablehead{
\colhead{3CR} & \colhead{RA\tablenotemark{a}} & \colhead{Dec\tablenotemark{a}} & \colhead{RA\tablenotemark{b}} & \colhead{Dec\tablenotemark{b}} & \colhead{Redshift} & \colhead{Filter} & \colhead{Rest-frame} & \colhead{Exposure} & \colhead{Dithers} & \colhead{Estimated Line} \\[0 cm]
& \colhead{(J2000)} & \colhead{(J2000)} & \colhead{(J2000)} & \colhead{(J2000)} & & & \colhead{$\lambda$ Range (\AA)} & \colhead{(s)} & & \colhead{Contamination\tablenotemark{c}}\\[0cm]
\colhead{(1)} & \colhead{(2)} & \colhead{(3)} & \colhead{(4)} & \colhead{(5)} & \colhead{(6)} & \colhead{(7)} & \colhead{(8)} & \colhead{(9)} & \colhead{(10)} & \colhead{(11)} 
}
\startdata
265	& 11h45m31.87s	& +31d33m43.4s	& 11h45m29.0s & +31d33m47s & 0.811 & F160W & 7805 - 9997 & 2559 & 4   & $\simlt 0.06$ mag ([SIII]$\lambda9533$)\\               \\ 
184	& 07h39m24.47s	& +70d23m10.9s	& 07h39m24.2s & +70d23m27s & 0.994 & F160W & 7089 - 9080 & 5134 & 6   & $\simlt 0.01$ mag\\                                     \\ 
280	& 12h56m57.49s	& +47d20m20.2s	& 12h56m57.7s & +47d20m21s & 0.996 & F160W & 7082 - 9071 & 5134 & 6   & $\simlt 0.01$ mag\\                                     \\ 
368	& 18h05m06.36s	& +11d01m32.6s	& 18h05m06.4s & +11d01m33s & 1.131 & F165M & 7293 - 8218 & 10268 & 12 & $\simlt 0.01$ mag\\                                     \\ 
267	& 11h49m56.53s	& +12d47m19.0s	& 11h49m56.6s & +12d47m19s & 1.140 & F165M & 7263 - 8183 & 10268 & 12 & $\simlt 0.01$ mag\\                                     \\ 
210	& 08h58m09.96s	& +27d50m51.6s  & 08h58m10.0s & +27d50m54s & 1.169 & F165M & 7166 - 8074 & 10268 & 12 & $\simlt 0.01$ mag\\                                     \\ 
65	& 02h23m43.19s	& +40d00m52.5s  & 02h23m43.4s & +40d00m53s & 1.176 & F165M & 7142 - 8048 & 7701 & 9   & $\simlt 0.01$ mag\\                                     \\ 
266	& 11h45m43.37s	& +49d46m08.2s	& 11h45m43.4s & +49d46m08s & 1.275 & F165M & 6832 - 7698 & 8727 & 13  & $\simlt 0.01$ mag\\                                     \\ 
470	& 23h58m35.34s	& +44d04m38.9s  & 23h58m36.0s\tablenotemark{d} & +44d04m45s\tablenotemark{d} & 1.653 & F160W & 5328 - 6824 & 7701 & 9   & $\approx 0.20$ mag (H$\alpha$)\\                        \\ 
239	& 10h11m45.42s  & +46d28m19.7s  & 10h11m45.3s & +46d28m20s & 1.781 & F160W & 5083 - 6510 & 7701 & 9   & $\approx 0.07$ mag ([OIII]$\lambda\lambda4959,5007$)\\  \\ 
256	& 11h20m43.05s	& +23d27m55.0s	& 11h20m43.0s & +23d27m54s & 1.819 & F165M & 5513 - 6212 & 7701 & 9 &$\simlt 0.01$ mag\\
&&&&&& F160W & 5014 - 6422 & 5134	& 6  & \\                                      \\ 

\enddata
\tablenotetext{a}{Radio source coordinates (J2000) from \citet*{DBB+96}}
\tablenotetext{b}{NICMOS galaxy coordinates (J2000) as measured from image WCS, subject to absolute errors of $\approx 1\arcsec$}
\tablenotetext{c}{Calculated assuming the composite spectrum of \citet{MCC93}}
\tablenotetext{d}{Assuming the middle source in the `trio' is the radio galaxy.  See the cutout in Figure 1}
\end{deluxetable}

\clearpage

\begin{deluxetable}{llrrrr}
\tabletypesize{\tiny}
\tablewidth{0pt}
\tablecaption{Galaxy Photometry\tablenotemark{a}}\label{magtab}
\tablehead{
\colhead{Object} & \colhead{Filter} & \colhead{$H_{160}$\tablenotemark{b}} & \colhead{$H_{160}$\tablenotemark{c}} & \colhead{$H_{160}$\tablenotemark{d}} & \colhead{$H_{160}$\tablenotemark{e}} \\[0cm]
&&\colhead{(Aperture Data)}&\colhead{(Aperture Model)}&\colhead{(Total Galaxy)}&\colhead{(Pt. Src.)}\\[0cm]
\colhead{(1)} & \colhead{(2)} & \colhead{(3)} & \colhead{(4)} & \colhead{(5)} & \colhead{(6)}
}
\startdata
3CR~265 & F160W & 18.67 & 18.70 & 18.74 & 22.84         \\
3CR~184 & F160W & 19.27 & 19.28 & 19.04 & $>$30.03      \\
3CR~280 & F160W & 19.53 & 19.51 & 19.49 & 23.03         \\
3CR~368 & F165M & 20.10 & 20.18 & 19.75 & $>$28.31      \\
3CR~267 & F165M & 20.19 & 20.22 & 20.07 & $>$25.12      \\
3CR~210 & F165M & 19.71 & 19.68 & 19.12 & $>$28.09      \\
3CR~65  & F165M & 19.26 & 19.43 & 19.35 & 22.75         \\
3CR~266 & F165M & 20.09 & 20.16 & 19.92 & $>$29.07      \\
3CR~470 & F160W & 21.15\tablenotemark{f} & \nodata & \nodata & \nodata   \\
3CR~239 & F160W & 19.88 & 19.87 & 19.68 & $>$27.19      \\
3CR~256 & F165M & 20.76 & \nodata & \nodata & \nodata   \\
\enddata
\tablenotetext{a}{All magnitudes are presented in the AB system}
\tablenotetext{b}{Measured in a 63.9 kpc diameter aperture, calculated in our adopted cosmology}
\tablenotetext{c}{Total aperture light of the best fit de Vaucouleurs model and point source (where applicable)}
\tablenotetext{d}{Total light in the best fit de Vaucouleurs model (no point source component)}
\tablenotetext{e}{Point source magnitudes.  Upper limits are 
derived from the maximal allowed (within 1$\sigma$ errors) point source component, 
and are shown for galaxies which were better fit by the pure galaxy model}
\tablenotetext{f}{Assuming the middle source in the 'trio' is the radio galaxy.  
See the cutout in Figure~\ref{TrioFig}}
\end{deluxetable}

\clearpage

\label{partab}
\begin{deluxetable}{rrrrrrrrrrrr}
\tabletypesize{\scriptsize}
\tablewidth{0pt}
\tablecaption{Derived NICMOS Galaxy Parameters\tablenotemark{a}}
\tablehead{
\colhead{Object} & \colhead{Outer Fit} & \colhead{$R_{e}$\tablenotemark{b}} & \colhead{$R_{e}$\tablenotemark{c}} & \colhead{$\langle I \rangle_{e}$\tablenotemark{d}} & \colhead{Ellip.\tablenotemark{e}} & \colhead{PA\tablenotemark{f}} & \colhead{Sersic Index} & \colhead{$k_{corr}\tablenotemark{g}$} & \colhead{M$_r$\tablenotemark{h}} & \colhead{M$_K$\tablenotemark{i}}  & \colhead{Stellar Mass\tablenotemark{j}}\\[0cm]
& \colhead{Radius ($\arcsec$)} & \colhead{($\arcsec$)} & \colhead{(kpc)} & \colhead{(mag/$\Box\arcsec$)}&&\colhead{(\deg E of N)} & \colhead{($n$)} & \colhead{(mag)} & \colhead{(mag)} & \colhead{(mag)}  &\colhead{($10^{11}$ \Msun)}\\[0cm]
\colhead{(1)} & \colhead{(2)} & \colhead{(3)} & \colhead{(4)} & \colhead{(5)} & \colhead{(6)} & \colhead{(7)} & \colhead{(8)} & \colhead{(9)} & \colhead{(10)} & \colhead{(11)} & \colhead{(12)}
}
\startdata
3CR~265 &    7.5 &   0.46 &   3.77 &  17.87 &  0.157 &    171 &    5.5 &   0.52 & -23.58 & -24.57 &   5.83 \\ 
3CR~184 &    3.8 &   1.10 &   9.46 &  19.59 &  0.227 &     82 &    2.6 &   0.37 & -23.86 & -24.85 &   6.33 \\ 
3CR~280 &    7.5 &   0.55 &   4.75 &  18.53 &  0.173 &     91 &    6.8 &   0.37 & -23.42 & -24.41 &   4.23 \\ 
3CR~368 &    7.5 &   1.54 &  13.62 &  20.75 &  0.157 &    171 &    5.7 &   0.32 & -23.49 & -24.45 &   4.01 \\ 
3CR~267 &    3.8 &   0.77 &   6.83 &  19.53 &  0.155 &     49 &    4.2 &   0.30 & -23.21 & -24.11 &   2.92 \\ 
3CR~210 &    7.5 &   1.93 &  17.14 &  20.50 &  0.367 &     10 &    2.8 &   0.27 & -24.24 & -25.14 &   7.54 \\ 
3CR~ 65 &    3.8 &   0.68 &   6.04 &  18.45 &  0.017 &      8 &    4.4 &   0.26 & -24.02 & -24.92 &   6.18 \\ 
3CR~266 &    3.8 &   1.09 &   9.82 &  19.83 &  0.324 &      5 &    2.6 &   0.19 & -23.70 & -24.57 &   4.20 \\ 
3CR~239 &    3.8 &   0.90 &   8.19 &  18.16 &  0.182 &    180 &    3.9 &  -0.18 & -24.98 & -25.56 &   5.90 \\ 
\enddata
\tablenotetext{a}{All magnitudes are presented in the AB system}
\tablenotetext{b}{Effective radius in arcseconds}
\tablenotetext{c}{Effective radius in kiloparsecs}
\tablenotetext{d}{Average surface-brightness within effective radius 
in rest-frame Gunn-$r$ 
magnitudes per square arcsecond}
\tablenotetext{e}{Ellipticity}
\tablenotetext{f}{Position angle}
\tablenotetext{g}{$k$-correction to rest-frame Gunn-$r$ magnitude using our model SED described in \S~\ref{LumMass}, $2.5 \textrm{log}_{10}(\langle f_{160} \rangle/\langle f_{r} \rangle$)}
\tablenotetext{h}{Absolute (total) Gunn-$r$ magnitudes}
\tablenotetext{i}{Absolute (total) rest-frame $K$ magnitudes}
\tablenotetext{j}{Stellar Mass as derived from the SED normalization 
to the NICMOS data}
\end{deluxetable}

\clearpage 

\begin{deluxetable}{rrrrrrrrrr}
\tabletypesize{\scriptsize}
\tablewidth{0pt}
\tablecaption{Derived WFPC2 Galaxy Parameters\tablenotemark{a}}\label{snaptab}
\tablehead{
\colhead{Object} & \colhead{Redshift} & \colhead{Filter} & \colhead{$R_{e}$\tablenotemark{b}} & \colhead{$R_{e}$\tablenotemark{c}} & \colhead{$\langle I \rangle_{e}$\tablenotemark{d}} & \colhead{Ellip.\tablenotemark{e}} & \colhead{PA\tablenotemark{f}} & \colhead{$k_{corr}\tablenotemark{g}$} & \colhead{M$_r$\tablenotemark{h}}\\[0cm]
&&& \colhead{($\arcsec$)} & \colhead{(kpc)} & \colhead{(mag/$\Box\arcsec$)}&&\colhead{(\deg E of N)}&\colhead{(mag)}&\colhead{(mag)}\\[0cm]
\colhead{(1)} & \colhead{(2)} & \colhead{(3)} & \colhead{(4)} & \colhead{(5)} & \colhead{(6)} & \colhead{(7)} & \colhead{(8)} & \colhead{(9)} & \colhead{(10)}
}
\startdata
3CR~ 35 &   0.03 & F702W &   5.87 &   3.85 &  21.07 &  0.281 &    112 &   0.08 & -20.42 \\ 
3CR~310 &   0.05 & F702W &  10.57 &  11.86 &  21.58 &  0.284 &     81 &   0.04 & -22.36 \\ 
3CR~403 &   0.06 & F702W &   8.35 &  10.27 &  20.86 &  0.320 &     34 &   0.02 & -22.77 \\ 
3CR~452 &   0.08 & F702W &   3.54 &   5.83 &  20.23 &  0.263 &    103 &  -0.02 & -22.17 \\ 
3CR~314 &   0.12 & F702W &   2.70 &   6.27 &  20.68 &  0.154 &     76 &  -0.10 & -21.87 \\ 
3CR~332 &   0.15 & F702W &   0.11 &   0.32 &  14.85 &  0.053 &     54 &  -0.16 & -21.24 \\ 
3CR~381 &   0.16 & F702W &   1.50 &   4.48 &  19.54 &  0.098 &     53 &  -0.18 & -22.28 \\ 
3CR~346 &   0.16 & F702W &   0.44 &   1.32 &  17.99 &  0.067 &    113 &  -0.19 & -21.18 \\ 
3CR~219 &   0.17 & F702W &   0.79 &   2.51 &  18.41 &  0.113 &    110 &  -0.21 & -22.16 \\ 
3CR~ 63 &   0.17 & F702W &   0.71 &   2.28 &  18.43 &  0.215 &     84 &  -0.21 & -21.92 \\ 
3CR~234 &   0.18 & F675W &   0.74 &   2.45 &  18.32 &  0.096 &    109 &  -0.30 & -22.20 \\ 
3CR~319 &   0.19 & F702W &   0.58 &   2.01 &  18.70 &  0.148 &    152 &  -0.25 & -21.38 \\ 
3CR~436 &   0.21 & F675W &   4.39 &  16.49 &  21.41 &  0.218 &     11 &  -0.36 & -23.25 \\ 
3CR~ 93 &   0.24 & F675W &   1.37 &   5.67 &  20.28 &  0.259 &    130 &  -0.43 & -22.06 \\ 
3CR~ 79 &   0.26 & F675W &   1.59 &   6.81 &  19.40 &  0.074 &     20 &  -0.45 & -23.34 \\ 
\enddata
\tablenotetext{a}{All magnitudes are presented in the AB system}
\tablenotetext{b}{Effective radius in arcseconds}
\tablenotetext{c}{Effective radius in kiloparsecs}
\tablenotetext{d}{Average surface-brightness within effective radius 
in rest-frame Gunn-$r$ 
magnitudes per square arcsecond}
\tablenotetext{e}{Ellipticity}
\tablenotetext{f}{Position angle}
\tablenotetext{g}{$k$-correction to rest-frame Gunn-$r$ magnitude using our model SED 
described in \S~\ref{LumMass}, $2.5 \textrm{log}_{10}(\langle f_{160} \rangle/\langle f_{r} \rangle$)}
\tablenotetext{h}{Absolute (total) Gunn-$r$ magnitudes}
\end{deluxetable}

\clearpage

\begin{deluxetable}{ll}
\tabletypesize{\scriptsize}
\tablewidth{0pt}
\tablecaption{Simulated Galaxy Parameters}\label{simtab}
\tablehead{
\colhead{Parameter} & \colhead{Values}}
\startdata
$R_e$ (pixels)\tablenotemark{a} & 3, 5, 10, 15, 30, 60 \\
$I_0$ (cts/s/pixel) & 100 \\
Point Source Strength \\(\% of total galaxy light) & 0, 1, 5, 10, 20 \\
Signal-to-Noise \\(per pixel at $R_e$) & 0.2, 1, 5, 15\\
\enddata
\tablenotetext{a}{These sizes correspond to effective radii of roughly 1 - 20 kpc 
at $z=1$ in our assumed cosmology, $(\Omega_{M},\Omega_{\Lambda}) = (0.3,0.7)$ and $H_0 = 65$}
\end{deluxetable}

\end{document}